\documentclass[11pt]{article}
\usepackage{jheppub}
\usepackage{amsmath,amssymb,amsfonts,graphicx}

\DeclareMathOperator{\arctanh}{arctanh}

\DeclareMathOperator{\Vol}{Vol}
\def\identity{{\rlap{1} \hskip 1.6pt \hbox{1}}}
\newcommand{\be}{\begin{equation}}
\newcommand{\ee}{\end{equation}}
\newcommand{\bea}{\begin{eqnarray}}
\newcommand{\eea}{\end{eqnarray}}
\newcommand{\beas}{\begin{eqnarray*}}
\newcommand{\eeas}{\end{eqnarray*}}
\newcommand{\ba}{\begin{array}}
\newcommand{\ea}{\end{array}}

\renewcommand*\d[2][]{%
	\mathrm{d}%
	\ifx\relax#1\relax\else
	\rule{-0.02em}{1.5ex}^{#1}\rule{0.08em}{0ex}\!
	\fi
	#2\,
}

\title{Holo-ween}

\author[]{Petar Simidzija, Mark Van Raamsdonk}

\affiliation[]{Department of Physics and Astronomy, University of British Columbia,\\
6224 Agricultural Road, Vancouver, B.C.\ V6T 1Z1, Canada.}

\emailAdd{psimidzija@phas.ubc.ca}
\emailAdd{mav@phas.ubc.ca}

\abstract{We argue that given holographic CFT${}_1$ in some state with a dual spacetime geometry $M$, and given some other holographic CFT${}_2$, we can find states of CFT${}_2$ whose dual geometries closely approximate arbitrarily large causal patches of $M$, provided that CFT${}_1$ and CFT${}_2$ can be non-trivially coupled at an interface. Our CFT${}_2$ states are ``dressed up as'' states of CFT${}_1$: they are obtained from the original CFT${}_1$ state by a regularized quench operator defined using a Euclidean path-integral with an interface between CFT${}_2$ and CFT${}_1$. Our results are consistent with the idea that the precise microscopic degrees of freedom and Hamiltonian of a holographic CFT are only important in fixing the asymptotic behavior of a dual spacetime, while the interior spacetime of a region spacelike separated from a boundary time slice is determined by more universal properties (such as entanglement structure) of the quantum state at this time slice. Our picture requires that low-energy gravitational theories related to CFTs that can be non-trivially coupled at an interface are part of the same non-perturbative theory of quantum gravity.}


\notoc

\begin{document}

\maketitle
\newpage
\parskip=10pt

\section{Introduction and motivation}

\begin{figure}
\centering
\includegraphics[width=80mm]{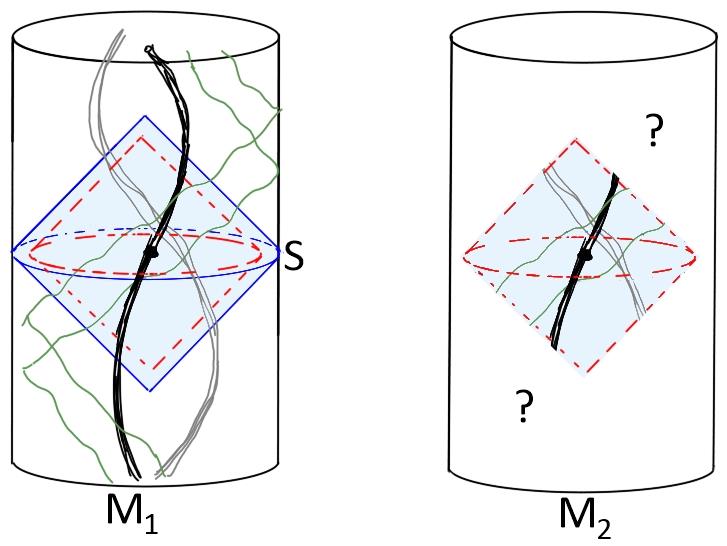}
\caption{Left: Wheeler-DeWitt patch $M_1^S$ (shaded blue region) of spacetime $M_1$ dual to CFT${}_1$ state $|\Psi_1 \rangle$ defined at boundary time slice $S$. Right: Geometry $M_2$ dual to CFT${}_2$ state $|\Psi_2 \rangle$ includes a region that approximates a large subset of $M_1^S$.}
\label{fig:Approx}
\end{figure}

Recent work has suggested a deep connection between the entanglement structure of quantum states of holographic systems and the geometry of the dual spacetimes they encode (see, for example \cite{Maldacena:2001kr,Ryu:2006bv,Swingle:2009bg,VanRaamsdonk:2009ar,VanRaamsdonk:2010pw,Maldacena:2013xja}, or \cite{VanRaamsdonk:2016exw} for a review). It is natural to ask how much information about the dual spacetime is captured by universal properties of the state such as entanglement structure and how much relates to more detailed properties of the CFT physics. In this paper, we will explore the extreme possibility that the precise microscopic degrees of freedom are actually unimportant, and that a suitably chosen state of any holographic CFT (or perhaps even some arbitrary QFT) can faithfully encode large regions of a spacetime dual to a completely different holographic CFT.\footnote{A particular motivation for this investigation is the work \cite{VanRaamsdonk:2018zws} which argued that states of a holographic CFT can be approximated by states of a large collection of non-interacting systems, each of which has almost no information about the global geometry. This suggested that the nature of these individual systems is not important, and that any quantum system with enough degrees of freedom, placed in an appropriate state, can faithfully encode the same gravitational physics. Another motivation is the recent work \cite{GP,AE,Almheiri:2019yqk}, which suggests that auxiliary radiation systems can encode the behind-the-horizon physics of late time black holes, regardless of what specific degrees of freedom the radiation system is built from.}

To be more precise, consider a state $|\Psi_1 \rangle$ of CFT${}_1$ dual to a spacetime geometry $M_1$. The CFT state defined on a particular boundary time slice $S$ encodes the ``Wheeler-DeWitt patch'' $M_1^S$ associated with $S$, defined as the union of spacelike bulk slices of $M_1$ ending on $S$, or equivalently the bulk domain of dependence of one of these slices.\footnote{Parts of the geometry outside this region would be affected by perturbations to the CFT${}_1$ Hamiltonian that happen before or after the time defined by $S$. Thus, these regions of the spacetime are not strictly encoded in the state $|\Psi_1 \rangle$ but require knowledge of $|\Psi_1 \rangle$ and the CFT${}_1$ Hamiltonian.}. We would like to argue that it is possible to find states of CFT${}_2$ whose dual geometries include regions which approximate arbitrarily large portions of $M_1^S$, as indicated in Figure \ref{fig:Approx}.

An immediate objection to this idea is that fields in the gravitational theory dual to a particular CFT (including Kaluza-Klein modes associated with the internal space) correspond to specific operators in the CFT. How can we hope to describe the same physics using a CFT with a completely different operator spectrum? To understand this, we recall that theories of gravity with different low-energy descriptions can be related to each other non-perturbatively, so we could have an interior spacetime region described by one semiclassical gravity theory appearing as a bubble in a spacetime whose asymptotic behavior is described by some very different semiclassical theory. In this case, the interior and exterior regions would be separated by some type of codimension-one brane (as in Figure \ref{fig:Euclidean}d).

Being able to encode CFT${}_1$ dual geometries using CFT${}_2$ states in this way requires that the low-energy theories of gravity which describe the asymptotic physics of these two different holographic CFTs must be part of the same non-perturbative theory of quantum gravity, which must include branes (i.e dynamical interfaces) that can connect regions with these two low-energy descriptions. Understanding which CFTs / gravitational theories are related in this way seems at first to be an impossible challenge, requiring complete knowledge of possible non-perturbative theories of quantum gravity. However we will now argue for a very simple sufficient condition: theories of gravity dual to different CFTs must be part of the same theory (which includes a non-perturbative interface brane), provided that the CFTs can be coupled non-trivially at an interface, such that energy and excitations can pass from one to the other.

\begin{figure}
\centering
\includegraphics[width=80mm]{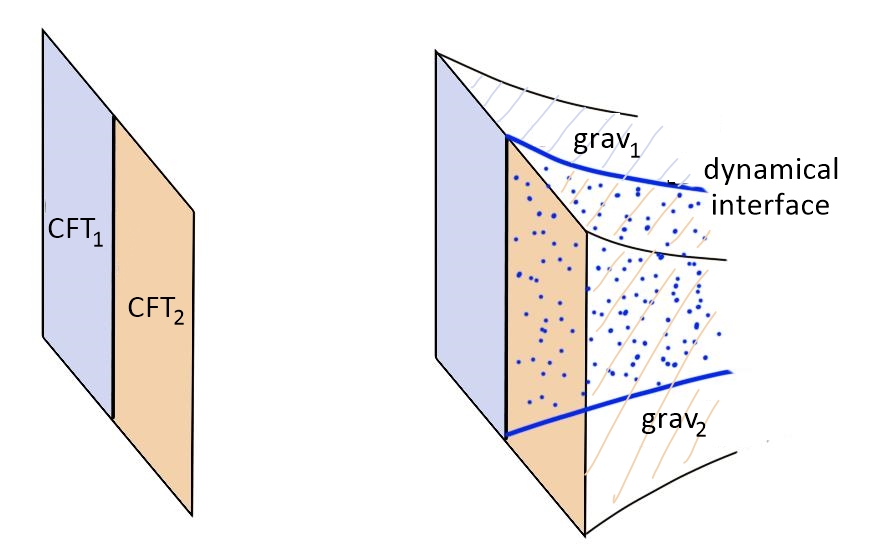}
\caption{Gravity interpretation of two holographic CFTs coupled non-trivially at an interface}
\label{fig:Interface}
\end{figure}

For suppose we have holographic CFTs joined along such an interface, and that we perturb one of the CFTs so that the perturbation is later detected in the other CFT. From the gravity perspective, we have an excitation in the region of spacetime associated with our first holographic CFT, and this leads to an excitation in the region of spacetime associated with the second holographic CFT. In this gravitational description, the implication is that the two regions of spacetime are joined somehow. In a gravitational theory, we expect that any such interface must be dynamical; this would be our desired codimension one ``brane'' separating regions with the two different low-energy descriptions (Figure \ref{fig:Interface}).

Given our assertion, it is very interesting to understand whether any two CFTs can be coupled non-trivially at an interface, and thus whether all low-energy gravitational theories are part of the same non-perturbative theory.\footnote{The recent works \cite{VanRaamsdonk:2018zws, McNamara:2019rup} have advocated for this picture.} In this paper we will set aside such questions; we will simply assume that CFT${}_1$ and CFT${}_2$ can be coupled non-trivially as described and focus on the question of how to construct states of CFT${}_2$ that faithfully encode the gravitational physics of a spacetime dual to some CFT${}_1$ state.

\subsubsection*{CFT${}_2$ disguised as CFT${}_1$}

\begin{figure}
    \centering
    \includegraphics[width=50mm]{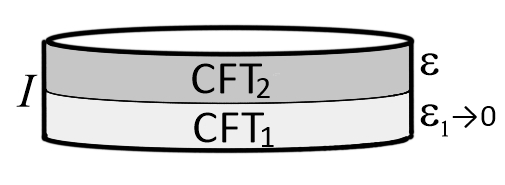}
    \caption{Euclidean path integral defining a quench operator $M_{{\cal I},\epsilon}$ mapping states of CFT${}_1$ to states of CFT${}_2$. Here, {\it I} represents a non-trivial conformal interface between the CFTs.}
    \label{fig:Quench}
\end{figure}

Supposing that $|\Psi_1 \rangle$ is a state of CFT${}_1$ with a good semiclassical description, our central idea is to define a state
\be
\label{Mapprox}
|\Psi^{\cal I}_2(\epsilon) \rangle = M_{{\cal I},\epsilon} |\Psi_1 \rangle
\ee
of CFT${}_2$ where $M_{{\cal I},\epsilon}$ is an operator defined by the Euclidean path-integral shown in Figure \ref{fig:Quench} (see section 2 for more details). In operator language, we have
\be
\label{Mop}
M_{{\cal I},\epsilon} = \lim_{\epsilon_1 \to 0} e^{-\epsilon H_2} {\hat Q}_{\cal I} e^{-\epsilon_1 H_1}
\ee
where the operator ${\hat Q}_{\cal I}$ performs a quench, modifying the Hamiltonian from that of CFT${}_1$ to that of CFT${}_1$ in a specific way associated with the chosen interface ${\cal I}$.\footnote{In general, there can be many possible interfaces between the same two theories.} This is a singular operation, resulting in a state of infinite energy since the UV structure of finite energy states for $H_2$ is different than the UV structure of finite energy states of $H_1$. The Euclidean time evolution $e^{-\epsilon H_2}$ removes the singular UV behavior, leaving us with a  state of finite energy with respect to $H_2$.

Since the operation $M_{{\cal I},\epsilon}$ corresponds to evolution for an infinitesimal Euclidean time with a local (but time-dependent) Hamiltonian, our hope is that if $\epsilon$ is sufficiently small, the new state $\Psi^{\cal I}_2(\epsilon)$ should have universal long-distance properties (e.g. entanglement structure) that closely resemble those of $|\Psi_1 \rangle$, such that $\Psi^{\cal I}_2(\epsilon)$ faithfully encodes a large portion of the spacetime dual to $|\Psi_1 \rangle$.

\begin{figure}
    \centering
    \includegraphics[width=150mm]{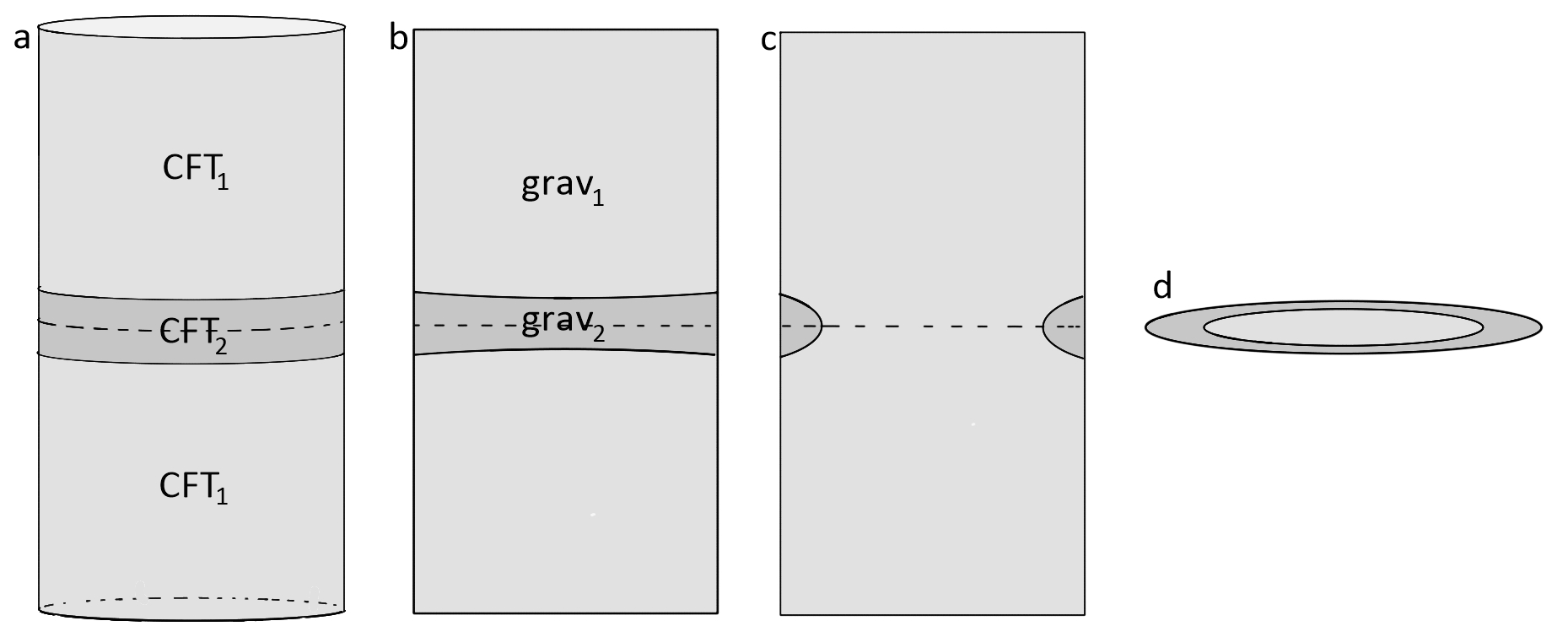}
    \caption{{\bf a)} CFT path integral geometry for $\langle \Psi_2|...|\Psi_2 \rangle$. {\bf b),c)} Possible topologies for the interface brane in the dual Euclidean solution (showing vertical cross section). {\bf d)} $t=0$ slice of c), providing initial data for the time-symmetric Lorentzian geometry dual to $|\Psi_2 \rangle$ in the case when solutions of the type c) have lowest action.}
    \label{fig:Euclidean}
\end{figure}

To check this, we would like to understand in detail the dual geometry to $|\Psi_2 \rangle$. For simplicity, we assume that the state $|\Psi_1 \rangle$ can also be described by a Euclidean path integral (possibly with sources). Then the Euclidean geometry dual to the CFT path integral of Figure \ref{fig:Euclidean}a used to compute observables in $|\Psi_2 \rangle$   will have an asymptotic region in which one strip is described by the semiclassical gravitational theory associated with CFT${}_2$ while the remainder is described by the semiclassical theory associated to CFT${}_1$. These two asymptotic regions must be separated by the brane associated with the chosen CFT interface.

Various topologies for the interface brane are possible, as shown in figure \ref{fig:Euclidean}b,c. The key idea that motivates the investigations in this paper is that if $\epsilon$ is sufficiently small, it is plausible that solutions with the connected-brane topology of figure \ref{fig:Euclidean}c will have the least action and dominate the gravitational path-integral. We see that a $\tau = 0$ slice of such a geometry (which gives the initial data for the Lorentzian geometry dual to $|\Psi^{\cal I}_2(\epsilon)$) contains a bubble of the geometry dual to CFT${}_1$, and the size of this bubble goes to infinity as $\epsilon \to 0$. We also expect that the interior geometry of this bubble approaches the geometry dual to $|\Psi_1 \rangle$. So we would have a state $|\Psi^{\cal I}_2(\epsilon) \rangle$ of CFT${}_2$ that describes an arbitrarily large bubble of the original spacetime dual to state $|\Psi_1 \rangle$ of CFT${}_1$.

\subsubsection*{Summary of results}

In this paper, we will check the expectations outlined in the previous paragraphs in the context of a simple model where the interface between CFT${}_1$ and CFT${}_2$ is associated with an interface brane with purely gravitational couplings. We show that the tension of this interface brane is directly related to the interface entropy\footnote{Via the ``folding trick'' the interface CFT can be understood as a boundary CFT whose bulk is the tensor product of CFT${}_1$ and (the mirror image of) CFT${}_2$ and whose boundary physics couples the two theories. The interface entropy is the boundary entropy for this theory.} (equation \ref{eq:loggkappa}) and that a finite interval of positive tensions can describe all possible interface entropies (equation \ref{eq:kappa_cond}).

In the context of our model, we consider the gravitational dual of states $|\Psi^{\cal I}_2(\epsilon) \rangle$ defined above, where the state $|\Psi_1 \rangle$ is taken at first to be the vacuum state of CFT${}_1$. We find the possible solutions as a function of the gravitational parameters $L^{AdS}_1$, $L^{AdS}_2$, the brane tension $\kappa$, and the ratio $S$ between the width of our strip in Euclidean time and the radius of the sphere on which the CFTs are defined.\footnote{Geometries of the type that we need have also been studied recently in \cite{Fu2019}, with a different motivation.} In the case of 1+1 dimensional CFTs, these correspond to the CFT central charges $c_1$,$c_2$, the boundary entropy $\log g$ and the parameter $\epsilon$ appearing in (\ref{Mop}).

For $c_2 > c_1$ and any boundary entropy, we find that the least action gravitational solution is always of the type {\bf c} in figure \ref{fig:Euclidean} (connected brane topology) for sufficiently small $\epsilon$. For larger values of brane tension / boundary entropy the bubble is behind black hole horizon.

For $c_1/3 < c_2 < c_1$ we find that the least action solution is of type {\bf c} for sufficiently large brane tension / interface entropy.

However, when $c_2 < c_1/3$ we find that there are no allowed solutions with topology of type {\bf c}. This would seem to contradict our original expectation, however, we will argue that the problem is that a transition directly from CFT${}_1$ to CFT${}_2$ with a much smaller central charge is too severe to preserve the desired properties of the original state. We proceed to consider more general interfaces which involve one or more intermediate CFTs and show that in this more general setup, it is always possible to choose a state of CFT${}_2$ such that the dual geometry contains an arbitrarily large causal patch of pure AdS geometry dual to the vacuum state of CFT${}_1$, in accord with our expectations.

In summary, within the context of our model, any CFT has states that can approximate arbitrarily large regions of the geometry dual to the vacuum state of any other CFT.

In the Lorentzian geometries dual to our CFT${}_2$ states, the interior picture is that we have a bubble of pure AdS spacetime that eventually collapses. The exterior picture is we have an asymptotically Schwarzschild-AdS spacetime with a spherically symmetric bubble wall that may initially be inside or outside the horizon but eventually falls in, collapsing to zero size at the singularity. This is shown in Figure \ref{fig:BHoptions_Lorentzian}.

In section \ref{sec:excited}, we argue that exactly the same construction (\ref{Mapprox},\ref{Mop}) works to approximate geometries dual to excited states of CFT${}_1$, both states perturbatively close to the vacuum and high-energy states dual to black hole geometries. In section \ref{sec:CFTstates}, we discuss some properties of the CFT${}_2$ states which encode CFT${}_1$ geometries, for example how much energy is needed to encode the physics of a bubble of size $r_0$ as a function of the central charges $c_1$ and $c_2$. We end with a discussion in section \ref{sec:discussion}, mentioning some directions for future work and offering a few more comments on the general picture.

Note added: after this manuscript was completed, we became aware of the work \cite{OTnew}, which also discusses the existence of CFT interfaces as a condition for the existence of interfaces between the corresponding gravitational theories.

\section{Interface CFTs}

Given two different CFTs, we can define a single theory on some spacetime where the local degrees of freedom on half of the space are those of CFT${}_1$ and the local degrees of freedom of the other half of the space are those of CFT${}_2$. At the interface, we can have some coupling between the CFTs on each side, possibly with some extra degrees of freedom living on the interface.\footnote{There are various ways we can imagine coupling two CFTs. If it is possible to understand the CFTs as arising from the IR limit of some UV lattice theories, we could consider the half-space version of each of these lattice theories (possible including additional degrees of freedom at the boundary) and then couple the two theories at the boundary in some way. The resulting theory should flow to a conformal theory in the IR with the local physics on either side of the interface described by CFT${}_1$ and CFT${}_2$. Similarly, if we can define the CFTs as the IR limits of UV theories with a field theory Lagrangian description, we can consider each of these UV theories on a half space and couple the fields in the two theories at the boundary. Again, the resulting theory should flow in the IR to a conformal theory with the local physics on either side of the interface described by CFT${}_1$ and CFT${}_2$.}

For our application, it will be important that the coupling between the CFTs is non-trivial, so that excitations/energy in one CFT can pass to the other CFT.

\begin{itemize}
\item
{\bf Definition:} We say that two CFTs are {\it compatible} if is possible to couple these CFTs to each other along an interface in a non-trivial way so that excitations/energy in one CFT can pass to the other CFT. More generally, we say $CFT_a$ and $CFT_b$ are in the same {\it compatibility class} if there is a sequence $(CFT_a, CFT_1, \dots, CFT_n, CFT_b)$ such that neighboring CFTs in the sequence are compatible.
\end{itemize}
It seems plausible that most CFTs are compatible. Generally speaking, when we have two ordinary physical systems, it is possible to couple them together in such a way that energy can be transferred from one to the other. It is also plausible that any two CFTs in the same compatibility class are compatible, since we could define an interface between them consisting of a set of layers involving $\{CFT_i\}$, and then consider the IR limit of this theory.

A useful way to think about an interface between two different CFTs is as a boundary condition for the CFT which is the tensor product between CFT${}_1$ and CFT${}_2$.\footnote{More precisely, we should take the mirror image of CFT${}_2$ if this is different that CFT${}_2$.} Here, we are using the so-called ``folding trick'' \cite{Oshikawa:1996dj}, where the half-space on which CFT${}_1$ is defined is identified with the half-space on which CFT${}_2$ is defined. From this perspective, we can say that CFT${}_1$ and CFT${}_2$ are compatible if there is a boundary condition for the product theory that couples the two bulk theories non-trivially.\footnote{Some boundary conditions for the tensor product theory correspond to choosing separate boundary conditions for CFT${}_1$ and CFT${}_2$ and taking the tensor product theory BCFT${}_1 \otimes$ BCFT${}_2$. These boundary conditions, or more general boundary conditions corresponding to linear combinations these, give rise to interfaces that we would describe as trivial, where excitations on one side of the interface simply reflect off the interface and are not transmitted to the other CFT.} This boundary condition can be a conformal boundary condition, or it can be an RG flow between conformal boundary conditions.

The allowed boundary conditions for the $CFT_1 \otimes CFT_2$ tensor product CFT are constrained by crossing symmetry and unitarity, and thus might in principle be classified by a bootstrap approach (see e.g. \cite{Liendo:2012hy}).

\subsection{Interface entropy}

For interface CFTs (where the interface theory is scale invariant) or for the UV or IR limits of interface RG flows, we can define an interface entropy to be the boundary entropy \cite{Affleck1991} of the associated boundary conformal field theory (see e.g. \cite{Azeyanagi:2007qj}). For 1+1 dimensional BCFTs, this boundary entropy is denoted $\log g$ and may be defined via the vacuum entanglement entropy for an interval of length $l$ that includes the boundary. Since the bulk central charge in the tensor produc theory is the sum of central charges for CFT${}_1$ and CFT${}_2$, we have
\be
\label{Interface}
S = {c_1 + c_2 \over 6} \log \left({2l \over \epsilon} \right) + \log g \; .
\ee
In higher dimensions, the boundary entropy can be defined via the vacuum entanglement entropy of a hemisphere centered on the boundary (see e.g. \cite{Estes:2014hka}).

\subsection{Approximating CFT states}

The central idea of this paper is to approximate a state of some holographic CFT${}_1$ by a state of another CFT${}_2$, preserving the entanglement structure and other ``universal'' properties of the state as much as possible at least above some UV length scale $\epsilon$.

If CFT${}_1$ and CFT${}_2$ were ordinary quantum systems with the same degrees of freedom (e.g. finite collections of qudits or harmonic oscillators), we could simply take $|\Psi_2 \rangle = |\Psi_1 \rangle$. However, for conformal field theories, it is not generally possible to identify the Hilbert spaces of the two systems, so we require a mapping from the states of one theory into the states of the other.

We can define such a mapping by making use of a Euclidean interface CFT, where the Euclidean versions of CFT${}_1$ and CFT${}_2$ are joined along an interface. We can define a map $M_{{\cal I},\epsilon}$ via its matrix elements:
\be
\langle \hat{\phi}_2 | M_{{\cal I},\epsilon} | \hat{\phi}_1 \rangle = \lim_{\epsilon_1 \to 0} \int_{\phi_1(- \epsilon_1) = \hat{\phi}_1}^{\phi_2(\epsilon) = \hat{\phi}_2}  \int [d \phi_i] e^{-S_{12}} \; .
\ee
Equivalently, we can write
\be
M_{{\cal I},\epsilon} = \lim_{\epsilon_1 \to 0} e^{-\epsilon H_2} {\hat Q}_{\cal I} e^{-\epsilon_1 H_1}
\ee
so that our CFT${}_2$ state is
\be
\label{eq:defstate}
|\Psi^{\cal I}_2(\epsilon) \rangle = M_{{\cal I},\epsilon} |\Psi_1 \rangle \; .
\ee
Here, the operator ${\hat Q}_{\cal I}$ performs a quench, modifying the Hamiltonian from that of CFT${}_1$ to that of CFT${}_2$. As we discussed in the introduction, this results in a state of infinite energy since the UV structure of finite energy states for $H_2$ is different than the UV structure of finite energy states of $H_1$. The Euclidean time evolution $e^{-\epsilon H_2}$ regulates the bad UV behavior, leaving us with a finite energy state of $H_2$. If $\epsilon$ is small, we expect that $e^{-\epsilon H_2}$ does not significantly change the entanglement structure, so the new state provides an approximation to the original state as far as the IR entanglement structure is concerned. The parameter $\epsilon_1$ which we have introduced to make the path integral expression well-defined, can safely be taken to zero.

Our goal will be to investigate the gravitational dual of the approximated state $|\Psi_2 \rangle$ and compare this with the gravitational dual of the original state $|\Psi_1 \rangle$. We take $|\Psi_1 \rangle$ to be a state constructed using a Euclidean path integral, possibly with sources for various operators. In this case, it will be possible to directly understand the gravitational duals of both $|\Psi_1 \rangle$ and $|\Psi_2 \rangle$ provided that we understand the gravitational physics associated with the CFT interface.

We emphasize that with our construction, we only expect the states $|\Psi_1 \rangle$ and $|\Psi_2 \rangle$ to be similar at the time when they are constructed. Since they evolve with different Hamiltonians, the state of CFT${}_2$ at some later time may bear little resemblance to the state of CFT${}_1$. However, it is expected that the CFT states encode the physics of a bulk region consisting of all bulk spatial slices anchored on the boundary slice at $t=0$ (or the domain of dependence region of one of these slices), as shown in Figure \ref{fig:Approx}. Thus, it is reasonable to hope that $|\Psi_2 (t=0)\rangle$ faithfully encodes the physics of a large portion of the interior of this spacetime region in the geometry dual to $|\Psi_1 \rangle$.

\section{Holographic interface CFTs}

We now consider the dual gravitational physics of the situation where we have two holographic CFTs coupled at a non-trivial interface. We have argued in the introduction that:
\begin{itemize}
\item
A physical system constructed from compatible CFTs joined by a non-trivial interface is dual to single gravitational theory with a {\it dynamical interface} separating two asymptotically AdS regions with different semiclassical descriptions.
\end{itemize}
We can think of the interface as a codimension one ``brane'' across which the low-energy description of the gravitational physics (cosmological constant, light fields, etc...) changes from that corresponding to the first CFT to that corresponding to the second CFT.\footnote{Such a brane could involve microscopic branes of the gravitational theory, or correspond to a region of spacetime in which the internal space makes a transition from one geometry to another. In some special cases, the brane can be described within the context of a single low-energy effective theory, for example where we have a scalar field making a transition between two possible extrema of its potential. This could correspond to an interface between CFTs for which CFT${}_2$ is the IR limit of an RG flow obtained by perturbing CFT${}_1$ with a relevant operator ${\cal O}$. In this case, the interface CFT can be understood as CFT${}_1$ with a spatially dependent coupling for  ${\cal O}$ that transitions from zero on one side of the interface to infinity on the other side of the interface.}

In this section, we introduce a simple holographic model where the gravitational dual includes different low-energy gravitational theories corresponding to the two CFTs and an interface brane that can connect regions described by these two low-energy theories.

\subsection{Bottom up model for holographic interface CFTs}

Gravity duals for interface CFTs have been considered various times in the literature. Precise gravitational solutions dual to the vacuum state of various supersymmetric interface CFTs have been worked out in \cite{Chiodaroli:2011fn, Chiodaroli:2012vc, DHoker:2007zhm, DHoker:2007hhe, Aharony:2011yc,Assel:2011xz} for example. In these ten-dimensional microscopic solutions, there is often no sharp interface brane, but rather a smooth ten-dimensional geometry with a region where the internal space makes a transition from one configuration to another. In a lower-dimensional description, the interface brane represents this transition region.

In this paper, we will employ a bottom-up approach, modelling the dual geometry for a $(D-1)$-dimensional interface CFT as a $D$ dimensional spacetime with two regions connected by an $(D-1)$-dimensional interface brane. This is similar to the bottom-up dual for BCFTs studied in \cite{Karch:2000gx,Takayanagi:2011zk,Fujita:2011fp} or the models in \cite{Azeyanagi:2007qj,Erdmenger:2014xya} used in the case of identical CFTs coupled at a defect. We take the gravitational action to be the Einstein action with a cosmological constant in each region related to the central charge of the corresponding CFT. In general, we can also have a bulk matter action in each region. At the interface, we have a Gibbons-Hawking-York boundary term associated with each side of the interface, plus an action for matter localized to the brane. This interface matter action can include couplings between fields on the interface and bulk fields. For our investigation, we will consider the simplest case where the interface brane is a constant-tension brane that couples only gravitationally to the bulk. The solutions we consider will then be purely gravitational and we can ignore the bulk matter actions. In summary, the relevant terms in the  action in the Euclidean case are
\begin{align}
\label{EucAc}
    \mathcal I=-\frac{1}{16\pi G_D}
    \Bigg[&
    \int_{\mathcal{M}_1}\d[D]{x}\sqrt{g_1}\left(R_1-2\Lambda_1\right)
    +\int_{\mathcal{M}_2}\d[D]{x}\sqrt{g_2}\left(R_2-2\Lambda_2\right)\notag\\
    &
    +
    2\int_\mathcal{S}\d[D-1]{y}\sqrt{h}(K_1-K_2)
    -2(D-2)\int_\mathcal{S}\d[D-1]{y}\sqrt{h} \kappa
    \Bigg],
\end{align}
where we have defined a tension parameter $\kappa\equiv 8\pi G_D T/(D-2)$ in terms of the tension $T$ and the cosmological constants $\Lambda_i$ are negative and related to the AdS lengths $L_i$ as $\Lambda_i = -(D-1)(D-2)/(2L_i)$. The equations of motion arising from varying this action are the Einstein equations with cosmological constant on each side of the interface, together with the second Israel junction condition \cite{Israel1966}
\be
\label{eq:JC2}
K_{1ab}-K_{2ab} = \kappa h_{ab}
\ee
where $K_{1ab}$ and $K_{2ab}$ represent the extrinsic curvatures for the interface, computed with the bulk metrics on either side, with the normal vector pointing from region 1 to region 2 in each case.

Note that we are also assuming the first junction condition, that there is a well-defined metric on the interface equal to the metric induced from the bulk metric on either side,
\be
\label{eq:JC1}
h_{1 ij} - h_{2 ij} = 0 \; .
\ee

In the next sections, following similar work in \cite{Takayanagi:2011zk} for boundary CFTs and \cite{Azeyanagi:2007qj} for defect CFTs, we establish a connection between the interface tension parameter $\kappa$ and the interface entropy for the corresponding interface CFT, by comparing the general result (\ref{Interface}) for interface entropy with a holographic calculation using the Ryu-Takayanagi formula \cite{Ryu2006b}.

\subsection{Planar interface solution}

We first derive the gravitational solution dual to the vacuum state of the interface CFT with a planar interface.

We consider two asymptotically AdS\textsubscript{3} spacetimes $(\mathcal M_1,g_1)$ and $(\mathcal M_2,g_2)$ with AdS lengths $L_1$ and $L_2$, separated by a domain wall of constant tension $T = \kappa/(8 \pi G)$. Near the AdS boundary, the metrics $g_1$ and $g_2$ can be written as
\begin{equation}
\label{eq:rhometric}
    d s_i^2 = d\rho_i^2+\frac{L_i^2}{y_i^2}\cosh^2\left(\frac{\rho_i}{L_i}\right)
    \left(-d t_i^2+d y_i^2\right),
\end{equation}
where $i\in\{1,2\}$, such that the AdS boundary occurs at $y_i=0$. A coordinate change $z_i=y_i/\cosh(\rho_i/L_i)$, $x_i=y_i\tanh(\rho_i/L_i)$ relates these coordinates to the usual Poincar\'e  coordinates
\begin{equation}
    d s_i^2 = \frac{L_i^2}{z_i^2}(d z_i^2-d t_i^2 +d x_i^2),
\end{equation}

\begin{figure}
    \centering
    \includegraphics[width=100mm]{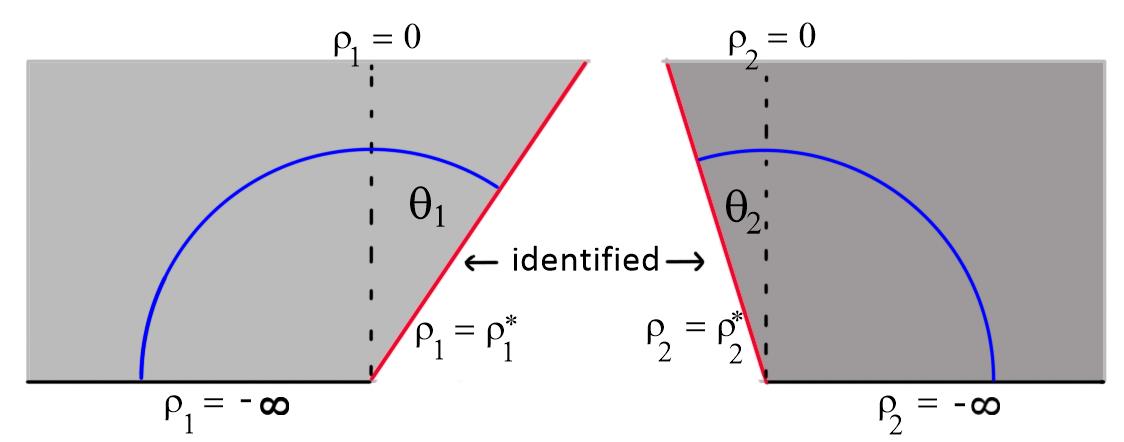}
    \caption{Gravity dual of an interface CFT in bottom up model with a constant tension-brane. The angles $\theta_1$ and $\theta_2$ in Poincare coordinates are determined by the AdS lengths $L_1$, $L_2$ and the brane tension $\kappa$. These are related to the CFT parameters $c_1$, $c_2$ and $\log g$. The RT surface for a spatial subsystem including points in each CFT within distance $l$ from the interface is shown in blue. }
    \label{fig:Planar}
\end{figure}

The location of the domain wall that separates these two regions can be determined by requiring that the junction conditions are satisfied. We make the ansatz that in the $(t_i,\rho_i,y_i)$ coordinates the domain wall is given by the embedding $\rho_i=\rho_i^*$, where we keep the part of spacetime $\mathcal M_i$ with $-\infty<\rho_i<\rho_i^*$, as shown in Figure \ref{fig:Planar}. Then the induced metric $h_{ab}$ on the domain wall $\mathcal S$, viewed as a surface in region $\mathcal M_i$, is
\begin{equation}
    d s_{\mathcal S,i}^2 = \frac{L_i^2}{y_i^2}\cosh^2\left(\frac{\rho_i^*}{L_i}\right)
    \left(-d t_i^2+d y_i^2\right).
\end{equation}
We see that the induced metric is independent of which embedding we use --- i.e. the first junction condition Eq.~\eqref{eq:JC1} is satisfied --- if we make the following identifications at the interface:
\begin{equation}
    y_1=y_2 \qquad t_1=t_2 \qquad L_1 \cosh\left(\frac{\rho_1^*}{L_1}\right)
    =
    L_2 \cosh\left(\frac{\rho_2^*}{L_2}\right).\label{eq:JC1_cond3}
\end{equation}

Next let us impose the second junction condition, Eq.~\eqref{eq:JC2}. It is straightforward to show that the extrinsic curvatures $K_{1ab}$ and $K_{2ab}$ of the domain wall embedded in regions $\mathcal M_1$ and $\mathcal M_2$ are given by
\begin{equation}
    K_{1ab} = \frac{1}{L_1}\tanh\left(\frac{\rho_1^*}{L_1}\right)h_{ab}\qquad
    K_{2ab} = -\frac{1}{L_2}\tanh\left(\frac{\rho_2^*}{L_2}\right)h_{ab},
\end{equation}
Then the second junction condition gives
\begin{equation}
\label{eq:JC2_cond1}
    \frac{1}{L_1}\tanh\left(\frac{\rho_1^*}{L_1}\right)
    +
    \frac{1}{L_2}\tanh\left(\frac{\rho_2^*}{L_2}\right)
    =
    \kappa,
\end{equation}
Equations ~\eqref{eq:JC1_cond3} and \eqref{eq:JC2_cond1} have the unique solution
\begin{align}
     \tanh {\rho_1^* \over L_1} &= \frac{1}{2} \left(\kappa L_1 + \frac{1}{\kappa L_1} -\frac{L_1}{\kappa L_2^2}\right),\label{eq:rho1}\\
     \tanh {\rho_2^* \over L_2} &= \frac{1}{2} \left(\kappa L_2 + \frac{1}{\kappa L_2} -\frac{L_2}{\kappa L_1^2}\right).\label{eq:rho2} \; .
\end{align}
Note that $\tanh {\rho_i^* \over L_i} = \sin(\theta_i)$, where $\theta_i$ is the angle by which the domain wall deviates from the normal to the AdS boundary in Poincar\'e coordinates.

\subsubsection*{Brane tension vs interface entropy}

Using our gravity solution, we can use the Ryu-Takayanagi (RT) formula \cite{Ryu2006b} to compute the interface entropy in terms of the bulk parameters and then compare the result with the CFT expression (\ref{Interface}) to come up with a relation between the brane tension and the interface entropy.

Starting from (\ref{Interface}), we note that the interface entropy can be expressed as
\be
\label{Sdiff}
\log g = S^{\cal I}_{[-l,l]} - {1 \over 2} S^1_{[-l,l]} - {1 \over 2} S^2_{[-l,l]}
\ee
where $S^{\cal I}$ is the vacuum entanglement entropy for the interval $[-l,l]$ in the interface CFT with interface at $x=0$, while $S^1$ and $S^2$ are the vacuum entanglement entropies for the same interval when CFT${}_1$ and CFT${}_2$ are defined on the whole real line.

The RT surface associated with the entropy of the interval $[-l,l]$ is built from extremal surfaces in the two regions which meet at the interface. The extremal surfaces are described in Poincar\'e coordinates as semicircular geodesic curves which intersect the AdS boundary orthogonally. In order that the two parts of the surface meet at the interface without a kink (which we require for an extremal area surface), it is also necessary that the curves meet the interface orthogonally on each side\footnote{To see this, we note that the Poincar\'e coordinate distance from the $x=0,z=0$ to the point where the geodesic intersects the interface is larger/smaller than $l$ if the intersection angle is larger/smaller than $\pi/2$. The intersection angles on the two sides sum to $\pi$ if there is no kink, so the only way for the distance along the interface to be the same on the two sides is to have both intersections be orthogonal.}. Thus, the RT surface is simply the the curve $t_i = 0, y_i = l$ (or $z^2 + x^2 = l^2$ in Poincar\'e coordinates), as shown in Figure \ref{fig:Planar}.

According to the RT formula, the entropy difference on the right side of (\ref{Sdiff}) is equal to $1/4G$ times the area of this $y_i = l$ surface (which covers $\rho_i = [-\infty,\rho_i^*]$) minus the area of the same surface $y_i = l$ for $\rho_i = [-\infty,0]$ in the pure AdS spacetimes corresponding to each side. From the metric (\ref{eq:rhometric}), we see that the area of a segment of the RT surface is just the coordinate length in the $\rho$ direction. Thus, we have simply
\be
\log g = {1 \over 4G} (\rho_1^* + \rho_2^*)
\ee
where $\rho_1^*$ and $\rho_2^*$ are given in (\ref{eq:rho1}). We can simplify the result somewhat to obtain
\be
\label{eq:loggkappa}
\log g = {L_1 + L_2 \over 4 G} \arctanh \left( {\kappa \over {1 \over L_1} + {1 \over L_2}} \right) + {L_1 - L_2 \over 4 G} \arctanh \left( { {1 \over L_1} - {1 \over L_2} \over \kappa} \right)
\ee
In the case $L_1 = L_2$, our result agrees with the earlier result \cite{Azeyanagi:2007qj} for the entropy in defect CFTs.

It is straightforward to check that for any positive $L_1,L_2$, we have a monotonic relation between $\log g$ and $\kappa$ in the interval
\begin{equation}\label{eq:kappa_cond}
    I = (\kappa_-,\kappa_+) \equiv \left(\left|{1 \over L_1}-{1 \over L_2}\right|,{1 \over L_1}+{1 \over L_2}\right).
\end{equation}
 that takes $(\kappa_-,\kappa_+)$ to $(-\infty, \infty)$. Thus, we have a bijection between CFT parameters $\{c_1>0$, $c_2>0$, $\log g\in\mathbb R\}$ and gravitational parameters $\{L_1>0$, $L_2>0$, $\kappa\in I\}$. 

It may be useful to note that the interface brane tension in holographic models has also been related \cite{Bachas:2020yxv} to the {\it transmission coefficient} \cite{Quella:2006de,Meineri:2019ycm}, a quantitative measure the extent to which excitations can pass from one CFT to another across a particular interface. 

\section{Gravity duals for approximated states}

\begin{figure}
    \centering
    \includegraphics[width = 50mm]{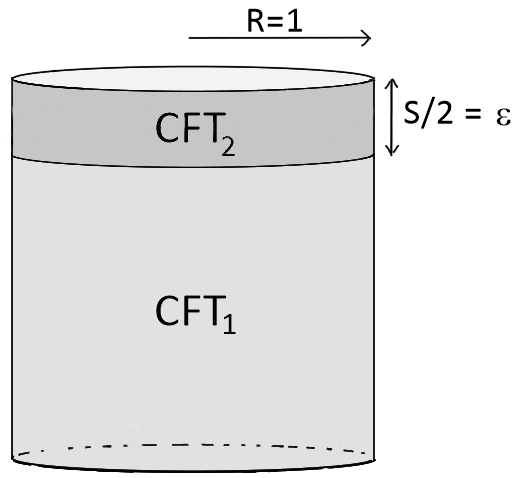}
    \caption{Euclidean path integral defining state $|\Psi_2(I, \epsilon) \rangle$ of CFT${}_2$ approximating the vacuum state of CFT${}_1$.}
    \label{fig:PIstate}
\end{figure}

In this section, we make use of the holographic model for interface CFTs outlined in the previous section to investigate the geometries dual to the states (\ref{eq:defstate}), where a state of holographic CFT${}_1$ is approximated by a state of holographic CFT${}_2$ by a Euclidean quench. In this section, we will focus on the case where the state $|\Psi_1 \rangle$ is the vacuum state of CFT${}_1$. This can be constructed via the path integral on a semi-infinite cylinder, so the state $|\Psi_2 \rangle$ arises from the Euclidean interface CFT path integral of figure \ref{fig:PIstate}. We follow the standard AdS/CFT recipe to understand the dual geometry to this state; the same type of geometries were considered recently in \cite{Fu2019}; some of our technical results overlap with results in that paper, but we present the solutions for completeness. First we consider the Euclidean geometry dual to the doubled path integral of figure \ref{fig:Euclidean}a. This will be a Euclidean time-reversal symmetric geometry whose $t=0$ slice provides the initial data for the Lorentzian geometry dual to state $|\Psi_2 \rangle$.

The boundary theory for the path integral in figure \ref{fig:Euclidean}a is a Euclidean CFT with two interfaces separated by some distance $S$ (which we will later take to be a small parameter $2 \epsilon$). These partition the Euclidean cylinder $\mathbb R\times S^{D-2}$ into three sections. Since the theories we are dealing with are scale-invariant, we can set the radius of the $S^{D-2}$ to be $R=1$ without loss of generality. The top and bottom sections are semi-infinite cylinders with a CFT of central charge $c_1$, while the middle section is a finite cylinder of height $S$ and contains a CFT of central charge $c_2$. We assume that both interfaces are the same, with defect entropy, $\log g\in \mathbb R$. We would like to find the possible bulk geometries corresponding to this setup.

As we have discussed, the dual geometries will have a dynamical interface brane separating regions in which the physics is described by the semiclassical gravitational theories dual to CFT${}_1$ and CFT${}_2$. In general, we expect two different possible topologies for this interface brane, as shown in Fig.~\ref{fig:Euclidean}b,c. The first type of geometry has two domain walls, with each wall ending on a single defect. The second type of geometry has a single domain wall running from one defect, into the bulk, and then ending on the other defect in a time symmetric manner. In some cases, solutions with each topology are possible for the same set of boundary data, and we will need to choose the one with least action.

We will assume that the bulk geometries preserve the  $SO(D-1)$ symmetry of the CFT setup. Then, assuming that the bulk solution is purely gravitational, the local geometry in each region must be Schwarzchild-AdS, which we describe with the usual metric
\begin{align}
\label{eq:metric}
    d s_i^2 = f_i(r)\d t^2+\frac{\d r^2}{f_i(r)}+r^2 d \Omega_{D-2}^2 \; .
\end{align}
Here, $i\in\{1,2\}$ denotes whether the asymptotic region of the geometry corresponds to CFT${}_1$ or CFT${}_2$, and $f_i(r)$ is defined as
\begin{equation}
    f_i(r)=1+\lambda_i r^2-\frac{\mu_i}{r^{D-3}} \; .
\end{equation}
The parameter $\mu_i$ is proportional to the mass of the corresponding Lorentzian black hole. The dual for the CFT${}_1$ vacuum state $|\Psi_1 \rangle$ corresponds to the full Euclidean AdS geometry, with $\lambda_1 = 1/L_1^2$ and $\mu_1 = 0$. For non-zero $S$, the Euclidean geometry at large $|t|$ should approach this pure AdS geometry, so we must have $\mu_1 = 0$. On the other hand, we can have $\mu_2$ either zero or non-zero, and we will find that these two possibilities correspond to the two cases in Fig.~\ref{fig:Euclidean}b,c. For this reason we refer to the geometry in Fig.~\ref{fig:Euclidean}b as the \textit{pure AdS phase}, and the geometry in Fig.~\ref{fig:Euclidean}c as the \textit{black hole phase}. The Lorentzian geometries corresponding to the black hole phase will involve a black hole at late times, while the Lorentzian geometries in the AdS phase will be pure AdS at the classical level.

\subsection{Interface trajectories}

All geometries that we describe are symmetric under time-reversal and can be constructed by gluing together patches of AdS / Schwarzschild AdS with boundaries described by interface trajectories which we describe either by $t_i(r)$ or by $(t_i(s), r(s))$ where $s$ corresponds to the proper distance along the interface in the $r-t$ directions. Since the coordinate $r$ corresponds to the physical radius of the $S^{D-2}$, the first junction condition implies that the $r$ coordinate of a point on the interface must be the same in the coordinates on each side. There is no such relation between the $t$ coordinates, so we need to provide $t_1(r)$ and $t_2(r)$ separately.

For AdS geoemtries, the upper half is a patch $\{r \in [0,\infty), t_1 \in [t_1(r), \infty)\}$ of pure AdS with $\lambda_1 = 1/L_1^2$ joined to a patch $\{r \in [0,\infty), t_2 \in [0, t_2(r)]\}$of pure AdS with $\lambda_2 = 1/L_2^2$. For black hole geometries, we have a patch $\{r \in [0,r_{min}], t_1 \in [0, \infty)\} \cup \{r \in [r_{min},\infty), t_1 \in [t_1(r), \infty)\}$ of pure AdS joined to a patch $\{r \in [r_{min},\infty), t_2 \in [0, t_2(r)]\}$ of AdS Schwarzschild geometry with $\lambda_2 = 1/L_2^2$ and some nonzero $\mu_2= \mu$. We will see that $\mu$ can be related to the CFT parameter $S$.

In either case, the range of the $t$ coordinate in the large $r$ asymptotic region of the middle section of the geometry is $\sqrt{\lambda_2} t_2 \in [-S/2,S/2]$, since this gives the appropriate ratio between the height and the radius of the central portion of the cylinder in the asymptotic region.

The precise form of the trajectories may be determined by solving the junction conditions. As we explain in appendix \ref{appendix:junctions}, the second junction condition
\begin{equation}
    K_{1ab}-K_{2ab}=\kappa h_{ab}.
\end{equation}
leads to the relation
\be\label{JC2_Schwarzschild}
f_1 {dt_1 \over ds} + f_2 {dt_2 \over ds} = \kappa r
\ee
while the definition of the proper length parameter $s$ (and the first junction condition that induced geometries on either side of the interface match) gives
\be\label{normalization}
f_i \left({dt_i \over ds}\right)^2 + {1 \over f_i}  \left({dr \over ds}\right)^2 = 1 \; .
\ee
Using (\ref{JC2_Schwarzschild}) and (\ref{normalization}) we find
\be
\label{req}
\left({dr \over ds}\right)^2 - V_{eff}(r) = 0 \qquad  \qquad V_{eff}(r) \equiv f_1 - \left({f_2 - f_1 - \kappa^2 r^2 \over 2 \kappa r} \right)^2 \; ,
\ee
and
\bea
\label{teq}
 {dt_1 \over dr} &=& {1 \over f_1 \sqrt{V_{eff}}} \left({1 \over 2 \kappa r}(f_1 - f_2) + {1 \over 2} \kappa r\right) \cr
 {dt_2 \over dr} &=& -{1 \over f_2 \sqrt{V_{eff}}} \left({1 \over 2 \kappa r}(f_2 - f_1) + {1 \over 2} \kappa r\right) \; .
\eea
The first equation tells us that $r(s)$ will be the trajectory of a particle with zero energy in a potential $-V_{eff}(r)$.\footnote{In the corresponding Lorentzian geometries, the trajectories will be governed by the potential $V_{eff}(r)$.}

\subsection{AdS solutions}

We first describe the pure AdS phase solutions. The results of this section are valid in any dimension.

From equation (\ref{req}), we find that for $\mu_1 = \mu_2 = 0$, $V_{eff} = 1 + A r^2$, where
\begin{equation}
\label{defA}
    A = {1 \over 4 \kappa^2}(\sqrt{\lambda_1}+\sqrt{\lambda_2} + \kappa)(\sqrt{\lambda_1}+\sqrt{\lambda_2} - \kappa)(\sqrt{\lambda_1}-\sqrt{\lambda_2} + \kappa)(\sqrt{\lambda_2}-\sqrt{\lambda_1} + \kappa) \; .
\end{equation}
From (\ref{eq:kappa_cond}), the range of allowed values of $\kappa$ is $\kappa \in (|\sqrt{\lambda_1} - \sqrt{\lambda_2}|,\sqrt{\lambda_1} + \sqrt{\lambda_2})$, we see that $A$ is always positive,\footnote{More precisely, we have $0 < A \le {\rm min}(\lambda_1, \lambda_2)$.} so $dr/ds$ cannot change sign and the solution to equation (\ref{req}) will be a monotonic function from $[s_0, \infty)$ to $[0,\infty)$. Choosing $s_0=0$, we have
\be
r(s) = {1 \over \sqrt{A}} \sinh \left(\sqrt{A} s \right) \; .
\ee
From (\ref{teq}), we find that $t_1(r)$ and $t_2(r)$ are monotonic, with the following signs:
\be
\ba{lll}
\lambda_2 < \lambda_1 : \qquad \qquad & \sqrt{\lambda_1} - \sqrt{\lambda_2} < \kappa < \sqrt{\lambda_1 - \lambda_2} &\qquad \qquad \dot{t}_1 > 0, \dot{t}_2 > 0 \cr
& \sqrt{\lambda_1 - \lambda_2} < \kappa < \sqrt{\lambda_1} + \sqrt{\lambda_2} &\qquad \qquad \dot{t}_1 > 0, \dot{t}_2 < 0 \cr
\lambda_2 > \lambda_1 :\qquad \qquad & \sqrt{\lambda_2} - \sqrt{\lambda_1} < \kappa < \sqrt{\lambda_2 - \lambda_1} &\qquad \qquad \dot{t}_1 < 0, \dot{t}_2 < 0 \cr
& \sqrt{\lambda_2 - \lambda_1} < \kappa < \sqrt{\lambda_1} + \sqrt{\lambda_2} &\qquad \qquad \dot{t}_1 > 0, \dot{t}_2 < 0 \; .
\ea
\ee
The trajectories $t_1(r)$ and $t_2(r)$ can be determined analytically by integrating (\ref{teq}). Taking $t_2=0$ to be the time-symmetric point in the geometry, we must have $t_2(r = \infty) = S/(2 \sqrt{\lambda_2})$. We can choose  $t_1(r= \infty) =0$. The result is
\bea
t_1(r) &=& {1 \over \sqrt{ \lambda_1}} \arctanh \left( \kappa^2 - \lambda_2 + \lambda_1 \over 2 \kappa  \sqrt{  \lambda_1( 1 + A r^2)} \right) \cr
t_2(r) &=& {S \over 2 \sqrt{\lambda_2}} + {1 \over \sqrt{ \lambda_2}} \arctanh \left( \lambda_1 - \lambda_2  - \kappa^2 \over 2 \kappa \sqrt{ \lambda_2 ( 1 + A r^2)} \right)
\eea

Below, it will be useful to have expressions for $\Delta t_i = t_i(\infty) - t_i(0)$, which we can write as
\be
\Delta t_1 = {1 \over \sqrt{\lambda_1}} \arctanh \left( \kappa^2 - \lambda_2 + \lambda_1 \over 2 \kappa  \sqrt{  \lambda_1} \right) \qquad \Delta t_2 = {1 \over \sqrt{ \lambda_2}} \arctanh \left( \lambda_1 - \lambda_2  - \kappa^2 \over 2 \kappa \sqrt{ \lambda_2} \right)
\label{eq:AdSDTs}
\ee
In order that the two domain walls asymptoting to $t_2 = \pm S/2$ do not intersect in the interior of the geometry, we have a constraint that $\Delta t_2 < S/2$. This is relevant only to the case $\lambda_2 < \lambda_1 ,\sqrt{\lambda_1} - \sqrt{\lambda_2} < \kappa < \sqrt{\lambda_1 - \lambda_2}$. Explicitly, we require
\be
S > {2 \over \sqrt{ \lambda_2}} \arctanh \left( \lambda_1 - \lambda_2  - \kappa^2 \over 2 \kappa \sqrt{ \lambda_2} \right)
\ee
For smaller values of $S$, no AdS solution exists.

\subsection{Black hole solutions}\label{sec:BHsolutions}

Let's now consider the solutions with $\mu_2 = \mu > 0$. In this case
\begin{equation}
\label{eq:Veff}
    V_\text{eff}(r) = A r^2+1+\frac{B}{r^{D-3}}-\frac{C}{r^{2D-4}},
\end{equation}
where $A > 0$ is defined in (\ref{defA}) and  $B, C$ are defined as
\begin{equation}
    B = -\mu \left(\frac{\lambda_1-\lambda_2+\kappa^2}{2\kappa^2}\right) \qquad \qquad
    C = \frac{\mu^2}{4\kappa^2} \; .
\end{equation}
Since $A$ and $C$ are positive, the potential $V_\text{eff}$ is positive for large $r$ and negative for small $r$ so must have a root for positive $r$. For $D \ge 3$ it can be shown \cite{Fu2019} that the root is unique, and we call this $r_0$. Domain wall trajectories that reach the asymptotic boundary of AdS reach a minimum radius at $r_0$ before returning to $r = \infty$. Thus, solutions for $\mu > 0$ are ``black hole'' solutions of the type shown in figure \ref{fig:Euclidean}c. For $D=3$, we have explicitly that
\begin{equation}
\label{defr0}
    r_0 = \frac{\mu}{\sqrt{2\sqrt{\mu^2 \kappa^2 \lambda_1 + \mu \kappa^2 (\lambda_2 - \lambda_1-\kappa^2) + \kappa^4} + \mu(\lambda_2 - \lambda_1 - \kappa^2) + 2 \kappa^2}} \qquad D = 3
\end{equation}

\subsubsection*{Euclidean AdS-Schwarzschild geometry}

Before proceeding, we recall a few useful facts about the Euclidean AdS/Schwazschild geometry in (\ref{eq:metric}). We have that $f_2(r) = 0$ at $r=r_H$, where $r_H$ is the horizon radius, determined by
\be
1 + \lambda_2 r_H^2 - {\mu \over r_H^{D-3}} = 0 \; .
\ee
For $D=3$, we have\footnote{For $0 < \mu < 1$, we have a conical singularity at $r=0$.}
\be
\label{eq:defrH}
r_H = {\sqrt{\mu - 1 \over \lambda_2}} \qquad \qquad D=3\; .
\ee
The geometry is smooth at $r = r_H$, provided that $t$ is taken to be periodic with period
\be
\label{defbeta}
\beta = {4 \pi \over f_2'(r_H)}
\ee
which reduces in $D=3$ to
\be
\beta = {2 \pi \over \sqrt{\lambda_2 (\mu - 1)}} \qquad \qquad D=3 \; .
\ee
For our solutions, the black hole region includes only a finite interval $t \in [-S/2,S/2]$ in the asymptotic region $r \to \infty$.

\begin{figure}
    \centering
    \includegraphics[width = 100mm]{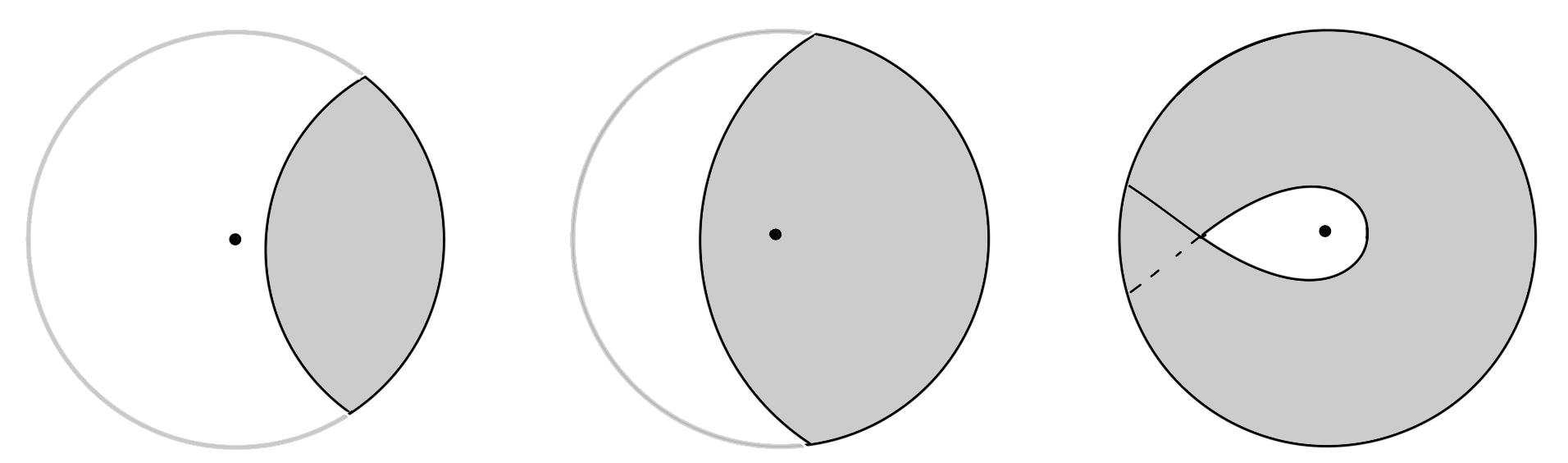}
    \caption{Possibilities for the Euclidean exterior geometry (shaded regions). Left: part of Euclidean AdS-Schwarzschild without a horizon.  Center: part of Euclidean AdS-Schwarzschild including the horizon (center point).  Right: solution without a horizon with $\Delta t_2 > \beta/2$.}
    \label{fig:BHoptions}
\end{figure}

We have two types of black hole solutions, characterized by whether or not the horizon is included in the spacetime region between the domain wall and the asymptotic AdS boundary (see figure \ref{fig:BHoptions}a,b). For the no-horizon solutions, $\dot{t}_2$ is positive near $r=r_0$, while for solutions with a horizon, $\dot{t}_2$ is negative for $r = r_0$. From (\ref{teq}), we find that the solution will include the Euclidean horizon\footnote{In the Lorentzian picture, this translates to the property that the domain wall is behind a black hole horizon at $t=0$. For the Euclidean solutions without a horizon, the domain wall is initially not behind a horizon but collapses to form a black hole at later times.} if
\be
f_2(r_0) - f_1(r_0) + \kappa^2 r_0^2 > 0 \; .
\ee
For $D=3$, this gives
\be
\label{eq:hashorizon}
\kappa > \sqrt{\lambda_1} \qquad \mu > 1 + {\lambda_2 \over \kappa^2 - \lambda_1} \qquad D = 3 \; .
\ee
We note that when the horizon is not included, our geometry can be a portion of a multiple cover of the Euclidean Schwarzschild solution, as shown in figure \ref{fig:BHoptions}c.

\subsubsection*{Interface trajectories}

The interface trajectories are again determined by integrating $\ref{teq}$. As for the AdS solutions, it will be useful to have expressions for $\Delta t_1$ and $\Delta t_2$. In this case, we have
\bea
\Delta t_1 &=& \int_{r_0}^\infty {dr \over f_1 \sqrt{V_{eff}}} \left({1 \over 2 \kappa r}(f_1 - f_2) + {1 \over 2} \kappa r\right) \cr
\Delta t_2 &=& -\int_{r_0}^\infty {dr \over f_2 \sqrt{V_{eff}}} \left({1 \over 2 \kappa r}(f_2 - f_1) + {1 \over 2} \kappa r\right) \; .
\eea
For $D=3$, we can give an explicit result in terms the of elliptic integral functions $\Pi$ and $K$
defined as
\begin{equation}
    \Pi(\nu,z) = \int_0^1 {dt \over (1 - \nu t^2) \sqrt{(1 - t^2)(1 - z^2 t^2)}} \qquad K(z) = \Pi(0,z)
\end{equation}
We have that
\begin{equation}
\label{eq:BHexact}
\Delta t_1 = z r_0 (K(z) + C_1 \Pi(\nu_1,z)) \qquad \Delta t_2 = {z r_0 \over 1 - \mu} (K(z) + C_2 \Pi(\nu_2,z)) \qquad \qquad D=3
\end{equation}
where
\begin{eqnarray}
z &=& {\mu \over \sqrt{\mu^2 + 4 r_0^4 \kappa^2 A}} \qquad
\nu_1 = {1 \over 1 + \lambda_1 r_0^2} \qquad
\nu_2 = {1 - \mu \over 1 - \mu + \lambda_2 r_0^2} \cr
C_1 &=&   {r_0^2 \nu_1 \over \mu}(\kappa^2 - \lambda_2 - \lambda_1(\mu-1)) \qquad
C_2 = {r_0^2 \nu_2 \over \mu } (-\kappa^2 + \lambda_1 + {\lambda_2 \over \mu-1})
\end{eqnarray}

For the black hole solutions without a horizon (figure \ref{fig:BHoptions} left and right), $t_2$ is 0 at the time-symmetric point and $S/(2 \sqrt{\lambda_2})$ at the boundary, so
\begin{equation}
\label{S1}
    S/2 = \sqrt{\lambda_2} \Delta t_2 \; .
\end{equation}
For the black hole solutions with a horizon (figure \ref{fig:BHoptions} center), $t_2$ decreases from $\beta/2$ at the time-symmetric point to $S/(2 \sqrt{\lambda_2})$ at the boundary, so
\begin{equation}
\label{S2}
    S/2 = \sqrt{\lambda_2} (\Delta t_2 + {\beta \over 2})  \; ,
\end{equation}
where $\beta$ is given in (\ref{defbeta}). In either case, we have a constraint that
\be
S \ge 0
\ee
in order that the brane does not intersect itself.

\subsubsection*{Existence of solutions}

\begin{figure}
    \centering
    \includegraphics[width=0.49\textwidth]{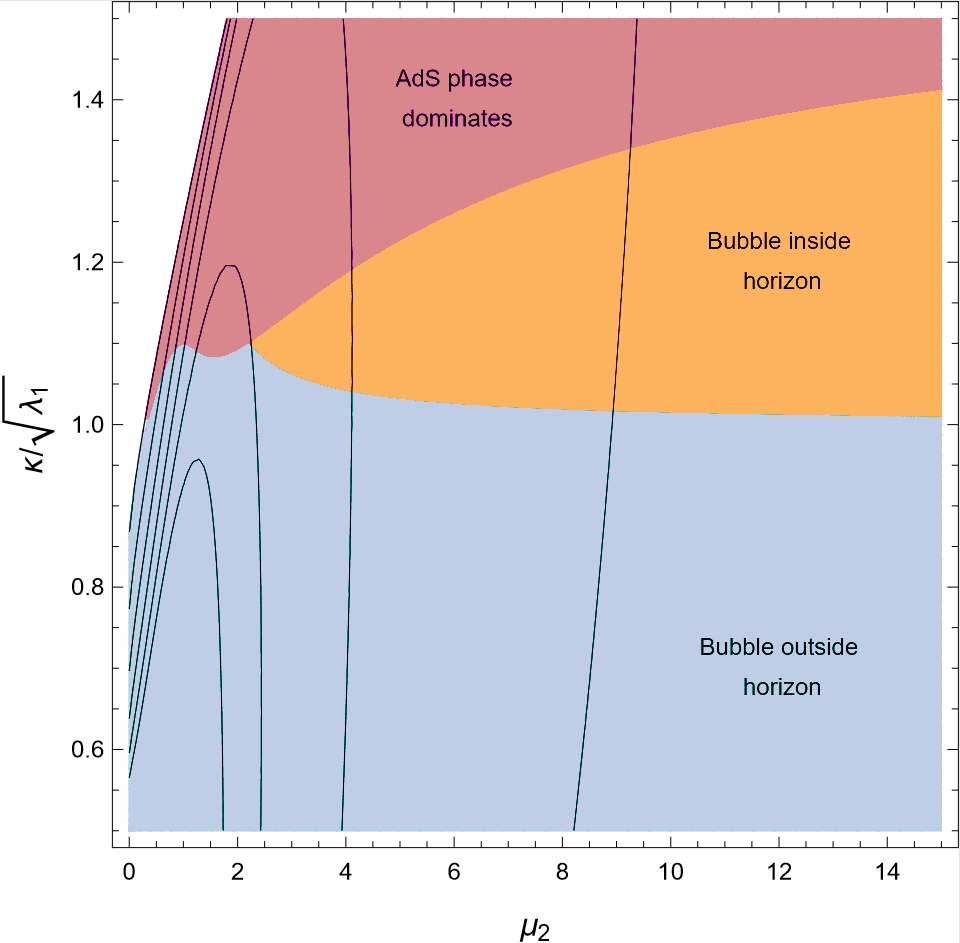}\hfill
    \includegraphics[width=0.49\textwidth]{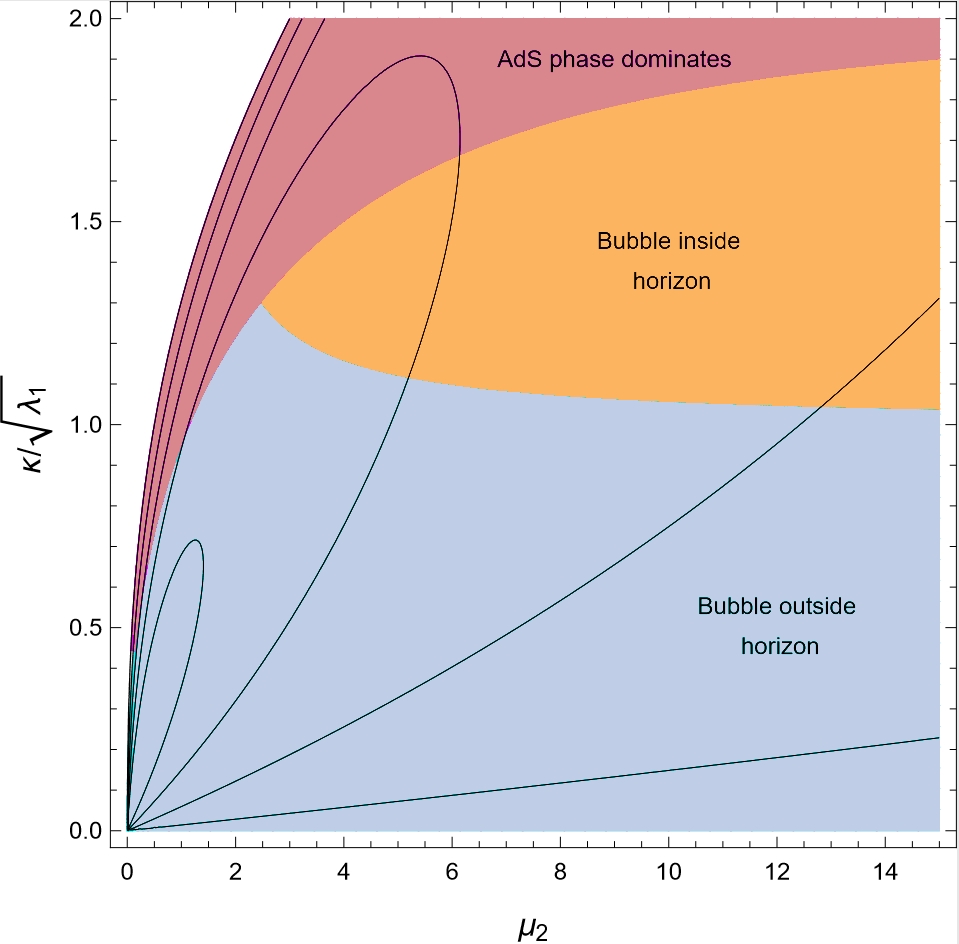}\\[\smallskipamount]
    \includegraphics[width=0.49\textwidth]{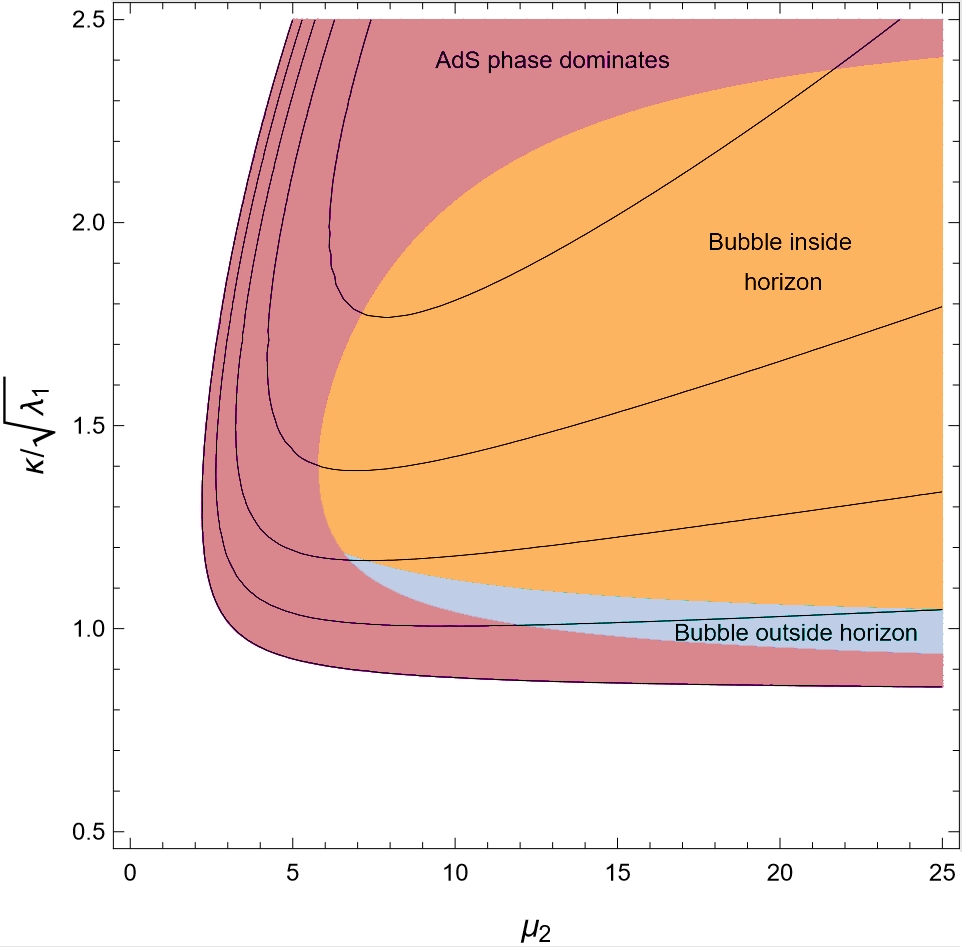}\hfill
    \includegraphics[width=0.49\textwidth]{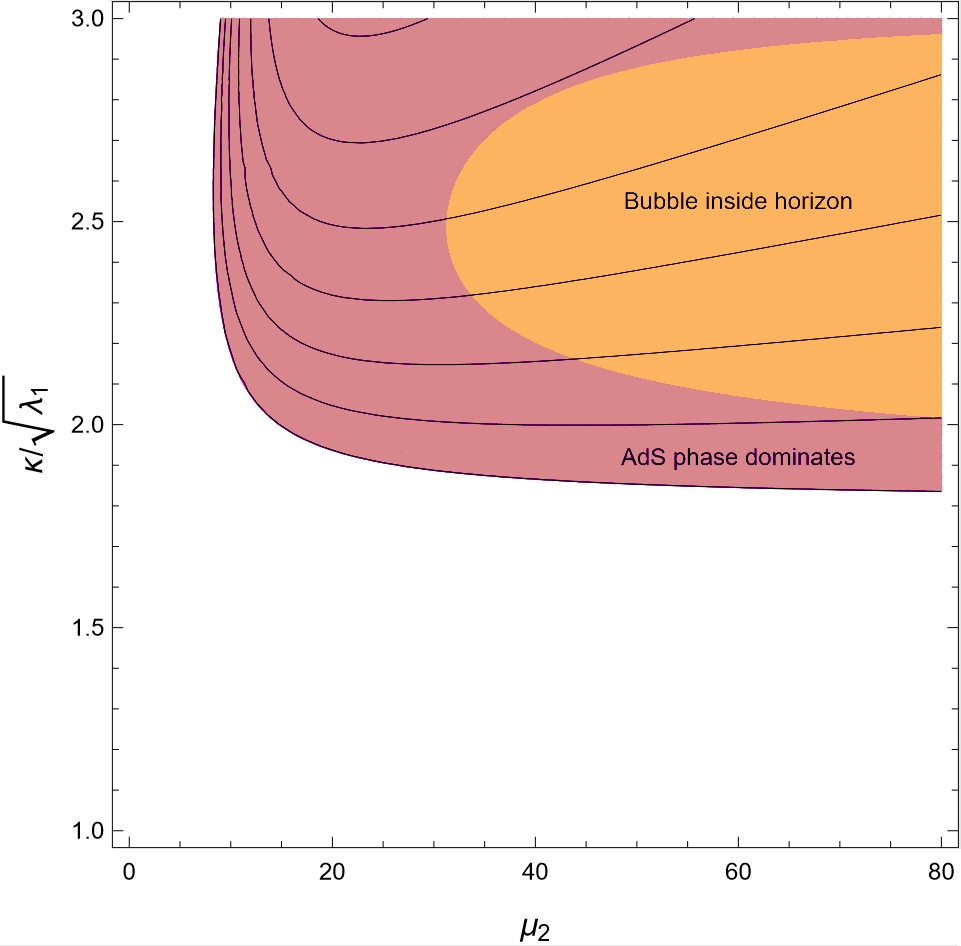}
    \caption{Phase plots with $S$ contours shown. $\sqrt{\lambda_2/\lambda_1}=1/2,1,3/2,2$ from top left to bottom right. The unshaded regions, corresponding to $S<0$, admit no black hole solutions.}
    \label{fig:mu_vs_kappa}
\end{figure}

The gravitational solutions depend on dimensionless parameters $\sqrt{\lambda_2/\lambda_1}$, $\kappa/\sqrt{\lambda_1}$, and $\mu$. The dimensionless quantity $S$ is a function of these parameters. The plots in figure \ref{fig:mu_vs_kappa} indicates the region of $\mu$-$\kappa$ parameter space for which black hole solutions of each type exist (for various values of $\sqrt{\lambda_2/\lambda_1}$) and also shows contour plots for the value of $S$. We find that black hole solutions satisfying the constraint $S>0$ exist only for $\sqrt{\lambda_2/\lambda_1} < 3$. For $\sqrt{\lambda_2/\lambda_1}$ in this range, we typically have solutions with horizons for larger values of $\kappa$ and solutions without horizons for smaller $\kappa$.

For fixed $\sqrt{\lambda_2/\lambda_1}$ and $\kappa/\sqrt{\lambda_1}$, we see that there can be two different values of $\mu$ that give rise to the same $S$. In our application, we are fixing the CFT parameter $S$, so we need to consider both solutions, along with the AdS solution with the same value of $S$ to determine which has least action.

\subsection{Comparing actions}

In cases where more than one gravity solution exists for the same CFT parameters $c_1$, $c_2$, $\log g$, $S$, the one that dominates the path integral and correctly computes CFT observables will be the one with least action. In Appendix \ref{app:actions}, we derive expressions for the action starting from the general expression (\ref{EucAc}). For $D=3$, the results are particularly simple. After removing a divergent term that is the same for each type of solution, we find that for the AdS geometries, the regulated action is
\begin{equation}
\label{acAdS}
    4G_3\mathcal{I}^{reg}_{AdS} = \Delta t_2 - \Delta t_1
\end{equation}
where $\Delta t_1$ and $\Delta t_2$ are given in (\ref{eq:AdSDTs}) while for black hole geometries, the regulated action is
\begin{equation}
\label{acBH}
    4G_3\mathcal{I}^{reg}_{BH} = (1 - \mu) {S \over 2 \sqrt{\lambda_2}} - \Delta t_1 \; ,
\end{equation}
where $\Delta t_1$ is given in (\ref{eq:BHexact}).

\subsubsection*{Phase diagrams for $D=3$}

Making use of the results in the previous section, we can compute the actions for the various solutions that are possible for fixed parameters $\lambda_1,\lambda_2,\kappa,S$. The diagrams in Figure \ref{fig:mu_vs_kappa} indicate which of the allowed black hole solutions have the least action compared with other black hole solutions or AdS solutions with the same parameter values.

\begin{figure}
    \centering
    \includegraphics[width=0.49\textwidth]{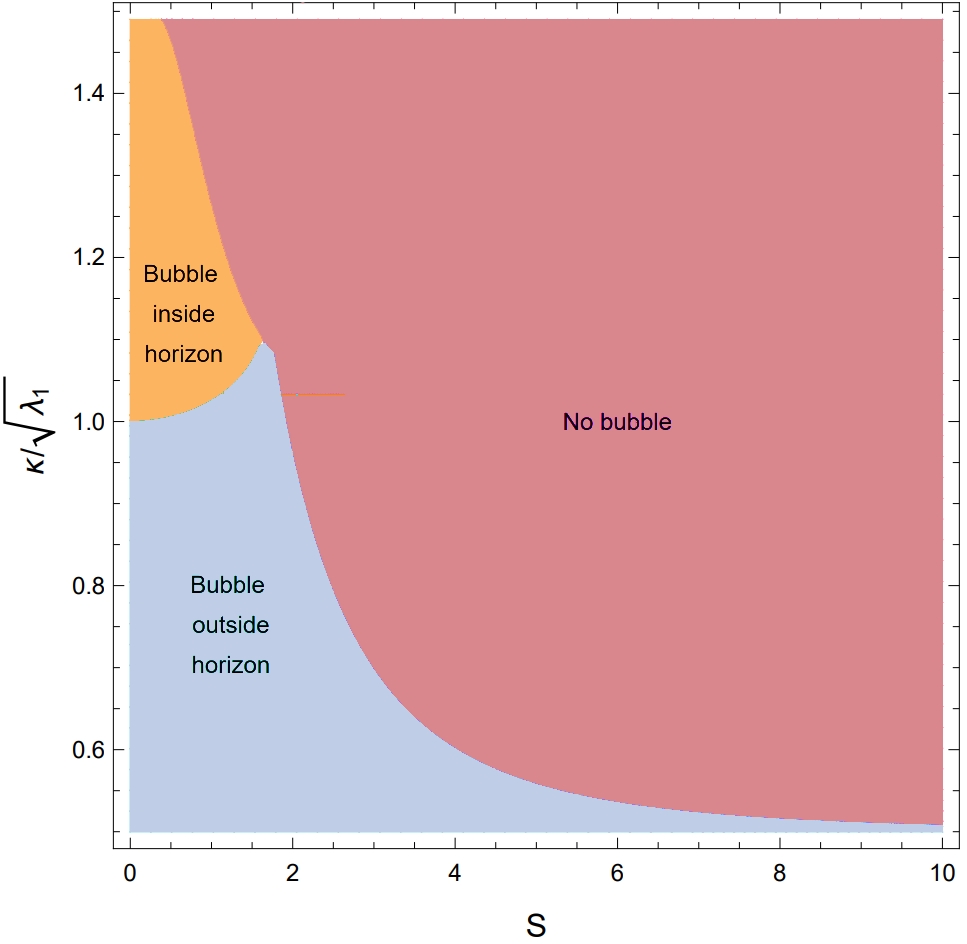}\hfill
    \includegraphics[width=0.49\textwidth]{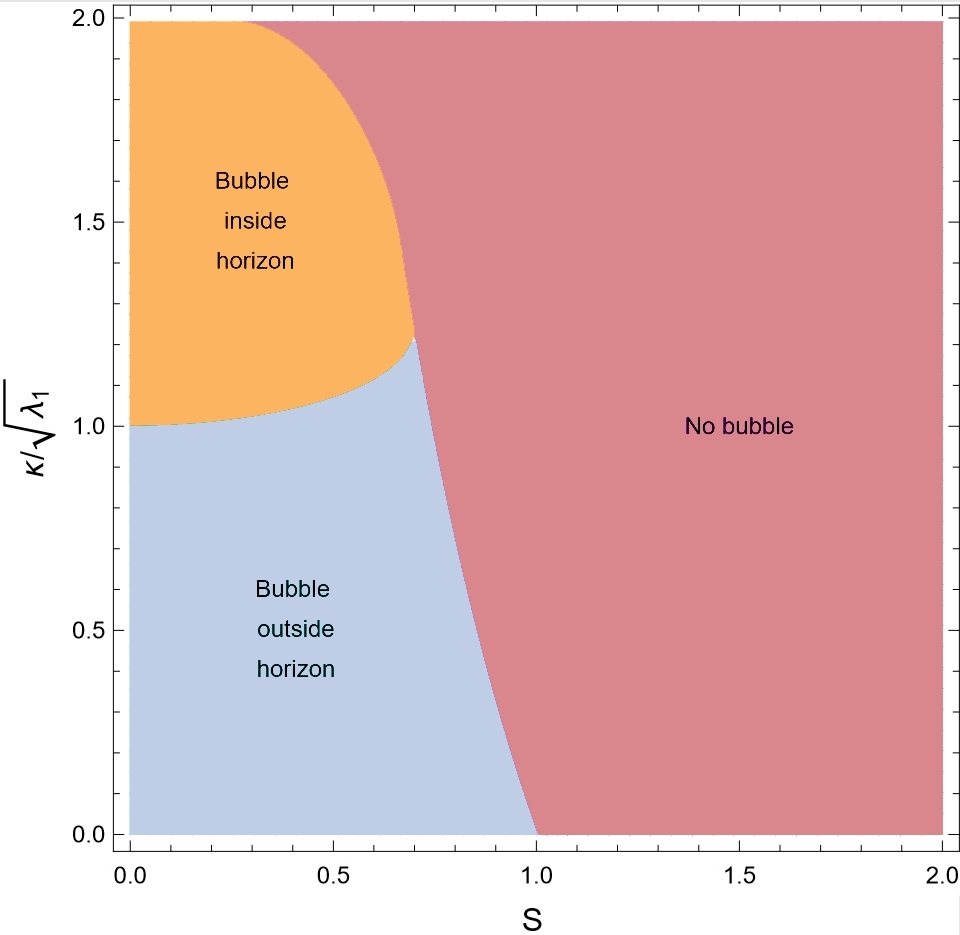}\\[\smallskipamount]
    \includegraphics[width=0.49\textwidth]{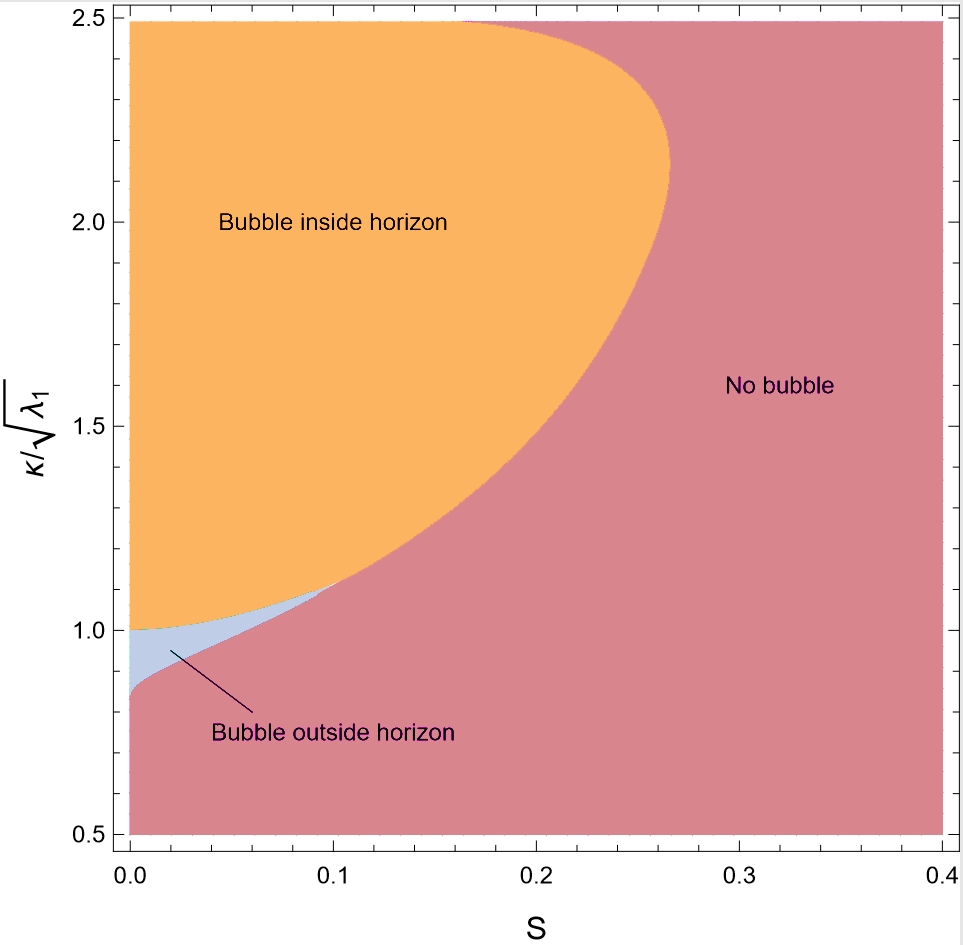}\hfill
    \includegraphics[width=0.49\textwidth]{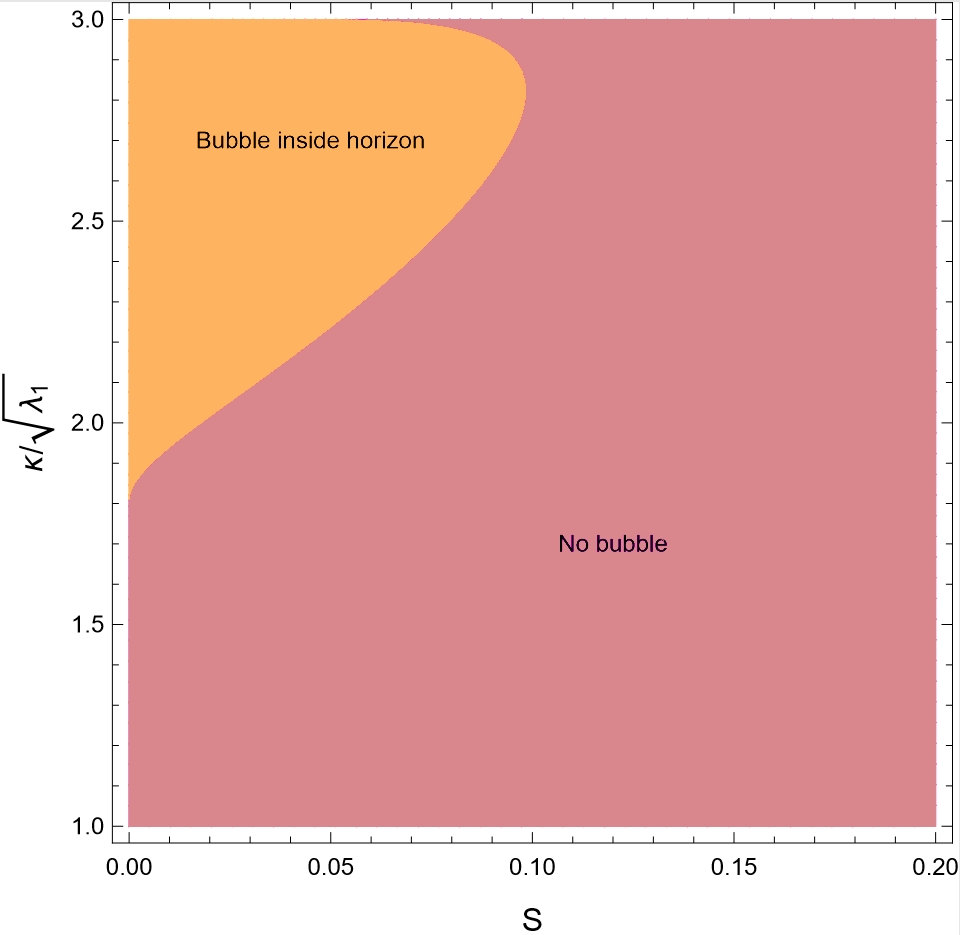}
    \caption{Phase plots showing least action gravity solution with $\sqrt{\lambda_2/\lambda_1}=1/2,1,3/2,2$ from top left to bottom right. For $\sqrt{\lambda_2/\lambda_1} > 3$, only AdS solutions exist.}
    \label{fig:S_vs_kappa}
\end{figure}

With this information, we can assemble phase diagrams showing the preferred solution as a function of $S$ and $\kappa$ for fixed $\sqrt{\lambda_2/\lambda_1}$. These are shown in figure \ref{fig:S_vs_kappa}.

\subsection{Small $S$}

For our application, we are mainly interested in the limit $S = \epsilon \to 0$, so that the Euclidean evolution with $H_2$ in (\ref{Mop}) does not significantly alter the long-distance properties of the CFT state. From the results in the previous section, we find that the least action solution is always a black hole solution with large $\mu$ when such a solution exists, or an AdS solution otherwise.

As we show in Appendix \ref{app:smallS}, in this large $\mu$ limit, we can write the relation between $S$ and $\mu$ explicitly as
\begin{equation}
\label{Smu}
     S = {1 \over \sqrt{\mu}} \left( 2 \pi \Theta(\hat{\kappa}-1) - \sqrt{\hat{\lambda}_2 \over \hat{\kappa}} K(z_\infty) - \sqrt{\hat{\lambda}_2 \over \hat{\kappa}} {\hat{\kappa} + 1 \over \hat{\kappa} -  1} \Pi \left(1 - {\hat{\lambda}_2 \over (\hat{\kappa}- 1)^2}, z_\infty \right)  \right),
\end{equation}
where
\begin{equation}
z_{\infty} = \sqrt{\hat{\lambda}_2 - (\hat{\kappa}-1)^2 \over 4 \hat{\kappa}} \qquad \hat{\lambda}_2 = {\lambda_2 \over \lambda_1} \qquad \hat{\kappa} = {\kappa \over \sqrt{\lambda}_1} \; ,
\end{equation}
so we see that $\mu \sim 1/S^2$ for small $S$. Legitimate black hole solutions exist provided that $S>0$. From (\ref{eq:hashorizon}) these include the Euclidean horizon if and only if $\kappa > \sqrt{\lambda}_1$.

Examining the expressions (\ref{acAdS}) and (\ref{acBH}) for the regulated actions, we see that the leading contribution at large $\mu$ to the action difference between a black hole solution and an AdS solution is
\begin{equation}
    S^{BH} - S^{AdS} = - {\mu S \over  2 \sqrt{\lambda_2}} + {\cal O} (\mu^{-{1 \over 2}})
\end{equation}
since the regulated action for the pure AdS solutions does not depend on $\mu$ and $\Delta t_1$ goes as $1/\sqrt{\mu}$ according to (\ref{eq:dt1app}). Thus, provided that a black hole solution exists (i.e. $S>0$), it will have lower action.

\begin{figure}
    \centering
    \includegraphics[width=100mm]{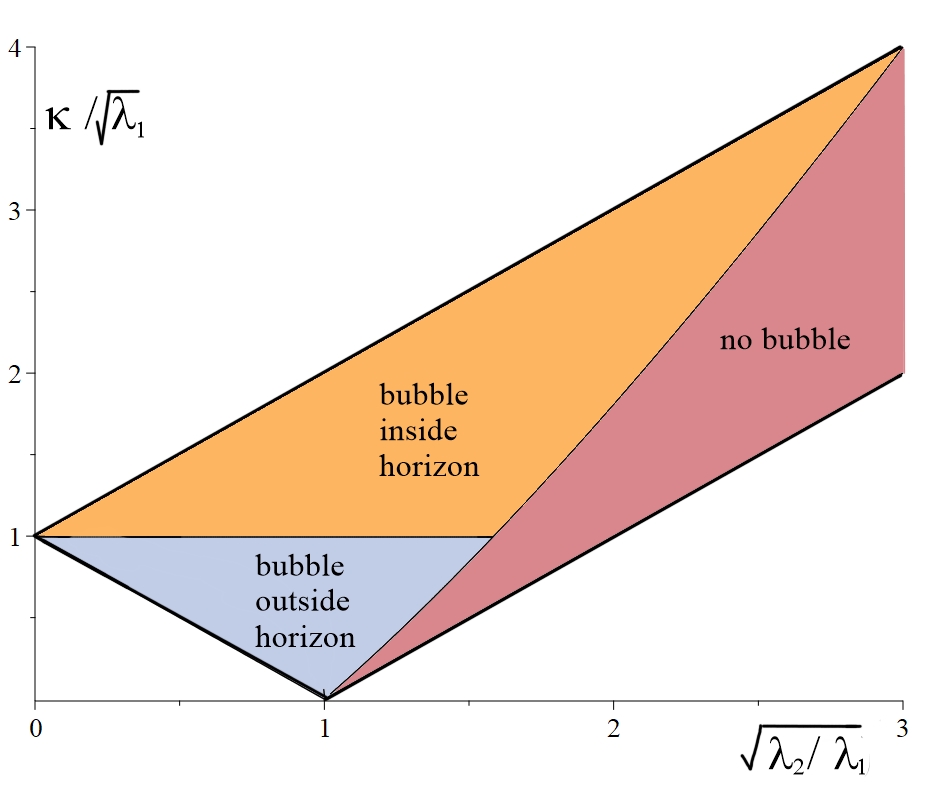}
    \caption{Phase diagram for small $S$.}
    \label{fig:smallS}
\end{figure}

Combining these results, we obtain the small $S$ phase diagram showing the least action solution as a function of $\sqrt{\lambda_2/\lambda_1}$ and $\kappa/\sqrt{\lambda_1}$ in Figure \ref{fig:smallS}.

At large $\mu$, the expression (\ref{defr0}) becomes
\be
r_0 = {\sqrt{\mu} \over \sqrt{\lambda_2 - (\sqrt{\lambda_1} - \kappa)^2}}
\ee
so we see that provided the black hole solution exists, the bubble of AdS vacuum can be made arbitrarily large by taking the limit of large $\mu$, or equivalently, small $\epsilon = S/2$.

\subsection{Summary of the results for small $S$}

We find that for $\lambda_2 < \lambda_1$ (i.e. if CFT${}_2$ has a larger central charge), the small $S$ solutions for any $\kappa$ are black hole solutions with a large bubble of the vacuum AdS geometry dual to CFT${}_1$. For $\kappa < \sqrt{\lambda_1 - \lambda_2}$, the solution does not include a Euclidean horizon. We will see below that the domain wall in the corresponding Lorentzian solutions is not initially behind a black hole horizon, but later collapses to form a black hole. For $\kappa > \sqrt{\lambda_1 - \lambda_2}$, the solution includes a Euclidean horizon and the domain wall in the Lorentzian solution is behind the black hole horizon at all times.

For $\lambda_2 > \lambda_1$ (i.e. when CFT${}_2$ has a smaller central charge), we have black hole solutions for some choice of $\kappa$ only when $\lambda_2 < 9 \lambda_1$ (or $c_2 > c_1/3$). For larger $\lambda_2$, all solutions are of AdS type, so apparently no CFT${}_2$ states of the form (\ref{eq:defstate}) encode the gravitational physics of the geometry dual to the CFT${}_1$ vacuum. On the other hand, we will see in section \ref{sec:multiple} that this goal can be achieved by considering CFT${}_2$ states using a path-integral with multiple interfaces.

\subsection{Lorentzian Solutions}

To find the Lorentzian geometries associated with our states, we use the $t=0$ slice of the Euclidean geometry as initial data for Lorentzian evolution. The resulting geometry is a portion of the maximally extended black hole geometry, truncated at some time-dependent radial location by the interface brane, with a portion of Lorentzian AdS glued on to the interior. Similar Lorentzian geometries have been discussed previously in \cite{Kraus:1999it, Fidkowski2003, Freivogel2005}.

\begin{figure}
    \centering
    \includegraphics[width = 100mm]{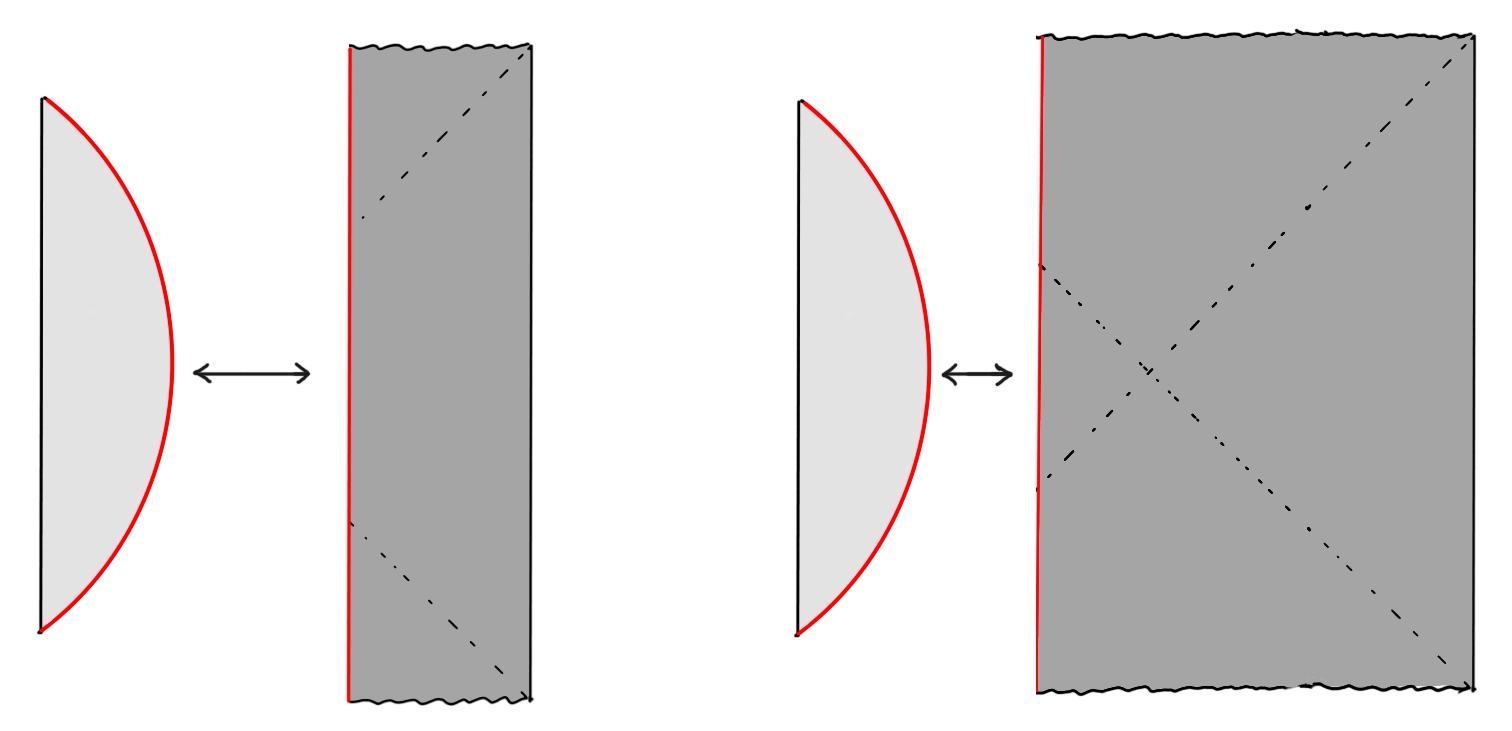}
    \caption{Schematic causal diagrams for Lorentzian solutions corresponding to Euclidean black hole phase solutions without (left) and with (right) a Euclidean horizon. In either case we have a collapsing bubble of pure AdS.}
    \label{fig:BHoptions_Lorentzian}
\end{figure}

The Lorentzian interface brane trajectory is described by analytically continued versions of the Euclidean equations (\ref{req}) and (\ref{teq}). The brane radius as a function of the proper time satisfies
\be
\label{req_Lorentzian}
\left({dr \over ds}\right)^2 + V_{eff}(r) = 0 \qquad  \qquad V_{eff}(r) \equiv f_1 - \left({f_2 - f_1 - \kappa^2 r^2 \over 2 \kappa r} \right)^2 \; .
\ee
Here, the sign of the potential is switched relative to the Euclidean case, so now the brane radius starts at $r=0$, expands to $r = r_0$ and then contracts again. From the interior point of view, this means that the size of our AdS bubble shrinks with time after $t=0$. From the exterior point of view, the interpretation is that the bubble emerges from the past singularity, grows, and then collapses to $r=0$ at the future singularity, as shown in Firgure \ref{fig:BHoptions_Lorentzian}. For solutions that include the Euclidean horizon, the bubble is behind the Lorentzian horizon at all times. For solutions without a Euclidean horizon, the bubble begins at the past singularity, emerges from the past horizon, falls back into the horizon, and then hits the future singularity. This description makes use of the worldvolume time on the brane. In terms of the exterior Schwarzchild time, the brane emerges from the horizon at $t=-\infty$ and falls back at $t = \infty$.

\subsection{Multi-interface solutions}
\label{sec:multiple}

\begin{figure}
    \centering
    \includegraphics[width = 100mm]{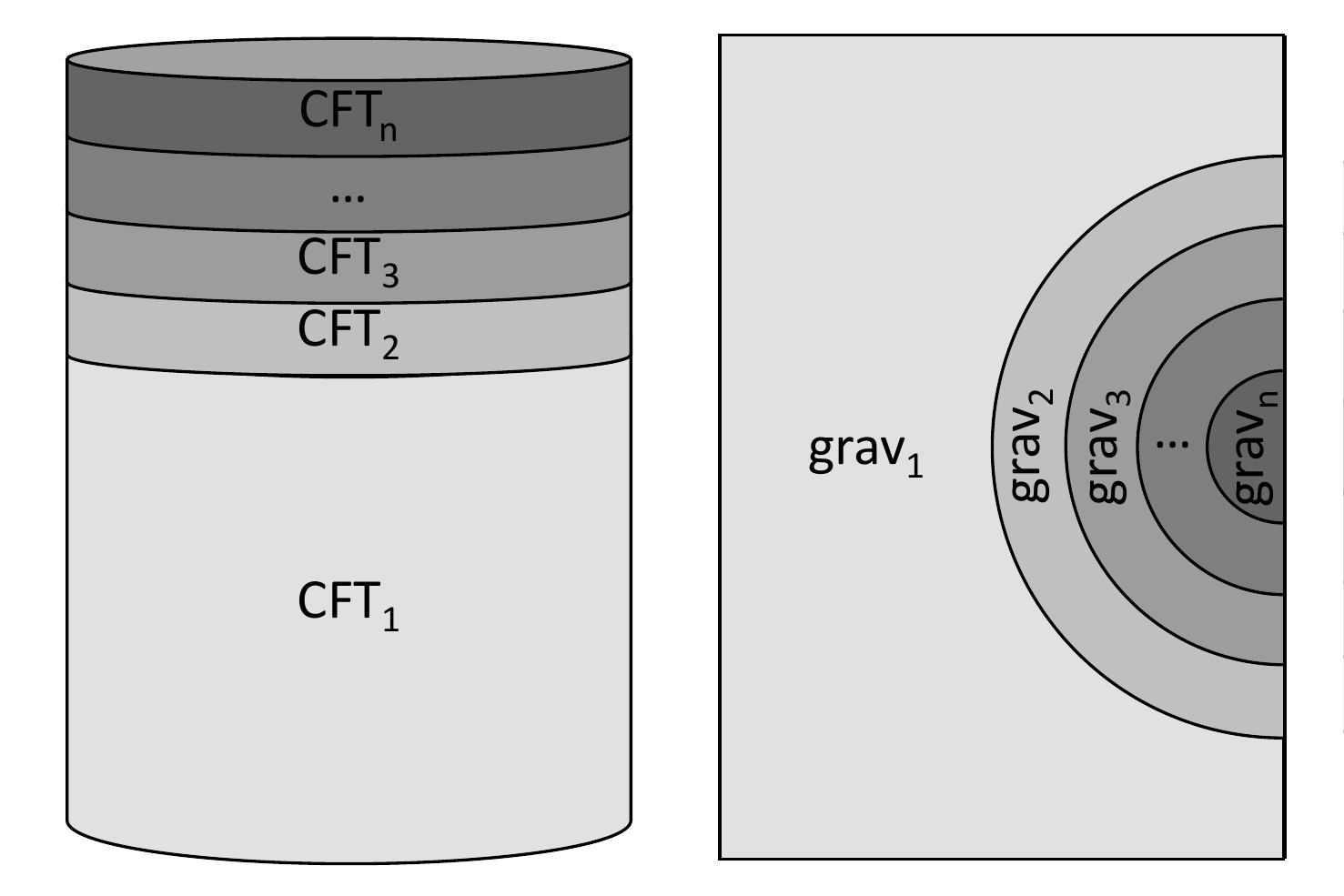}
    \caption{Left: construction of the state $|\Psi_n\rangle$ of CFT${}_n$ from the state $|\Psi_1\rangle$ of CFT${}_1$. Right: dominant gravitational dual in the case of large $\mu_i$, with $i\ge 2$.}
    \label{fig:multi_interface}
\end{figure}

In the previous section, we have found that when the central charge of CFT${}_2$ is too small, $c_2 < c_1/3$, we have only AdS solutions, suggesting that states $|\Psi_2 \rangle = e^{-\epsilon H_2} {\hat Q}_{\cal I} |\Psi_1 \rangle$ cannot provide an approximation to the CFT${}_1$ vacuum state that faithfully encodes the gravitational physics of a large region of the original pure AdS spacetime. On the other hand, we have found that states of a CFT${}_2$ with smaller central charge {\it can} faithfully encode large regions of the AdS spacetime dual to the CFT${}_2$ vacuum provided that $c_2/c_1$ is not too small. Below, we will argue that the construction (\ref{eq:defstate}) also works for excited states. This suggests the following procedure to construct a state of CFT${}_n$ with central charge smaller than $c_1/3$ that can properly encode the physics dual to the CFT${}_1$ vacuum:
\begin{itemize}
    \item start with the vacuum state of CFT${}_1$
    \item consider a sequence CFT${}_2$, CFT${}_3$, \dots CFT${}_n$ of CFTs with $c_{i+1}/c_i$ in the range $(1/3,1)$
    \item Construct a state $|\Psi_{i+1} \rangle$ of CFT${}_{i+1}$ from state $|\Psi_{i} \rangle$ of CFT${}_i$ by the map (\ref{eq:defstate}).
\end{itemize}
All together, we have a state of CFT${}_n$ defined from the CFT${}_1$ vacuum by a series of Euclidean quenches involving a series of intermediate CFTs (see left panel in Figure \ref{fig:multi_interface}):\footnote{A similar construction was considered in \cite{Fu2019} when constructing bag-of-gold spacetimes.}
\be
|\Psi_n \rangle = e^{-\epsilon_n H_n} \hat{Q}_{{\cal I}_{n \; (n-1)}}e^{-\epsilon H_{n-1}} \cdots e^{-\epsilon_2 H_2} \hat{Q}_{{\cal I}_{2 1}} |\Psi_1 \rangle \; .
\ee
One motivation for considering these states is that the quench operator ${\hat Q}_{\cal I}$ may necessarily result in too severe a modification to the original state when passing to a new CFT with much smaller central charge than the original CFT. By including more steps, we can try to make the transition more gradual, so that the final state can faithfully encode the original geometry.

We will consider a sequence of CFTs such that the ratio of central charges $c_{i+1}/c_{i}$ for neighboring CFTs is only slightly less than one, and the tension associated with the interface brane between these two CFTs is small. We take the width in Euclidean time of the strip with CFT${}_i$ to be $S_i/2$. On the gravity side, we have parameters $\lambda_i = 1/L_i^2$, $\mu_i$ ($i=1,\dots,n$), and $\kappa_i$ ($i=1,\dots,n-1$), where $\mu_1 = 0$.

For the interface between the $i$th and $(i+1)$st region, the trajectory is determined by
\be
\left({dr \over ds}\right)^2 = V_{eff}
\ee
with
\be
V_{eff} = 1 + \lambda_i r^2 - \mu_i - \left[ {(\lambda_{i+1} - \lambda_i - \kappa_{i}^2) \over 2 \kappa} r - {\mu_{i+1} - \mu_i \over 2 \kappa r}\right]^2 \; .
\ee
The $i$th region is bounded by the $(i-1)$st and $i$th domain walls, whose trajectories are described by $t_i^+(r)$, $t_i^-(r)$, satisfying
\bea
\label{tpmeq}
 {dt_i^- \over dr} &=& {1 \over f_i \sqrt{V_{eff}}} \left({1 \over 2 \kappa_i r}(f_i - f_{i+1}) + {1 \over 2} \kappa_i r\right) \cr
 {dt_i^+ \over dr} &=& -{1 \over f_i \sqrt{V_{eff}}} \left({1 \over 2 \kappa_{i-1} r}(f_i - f_{i-1}) + {1 \over 2} \kappa_{i-1} r\right) \; .
\eea
For a valid solution where the domain walls do not intersect, we require that $\Delta t_i^+ > \Delta t_i^-$, and $t_i^+(r) > t_i^-(r)$ for $r \ge (r_0)_{i+1}$.\footnote{Here, we will always be choosing $\kappa_i < \sqrt{\lambda_{i+1} - \lambda_i}$, so all $\dot{t}$s are positive.}

We would like to show that there is a valid solution with $\lambda_n/\lambda_1$ arbitrarily large ($c_n/c_1$ arbitrarily small). To see this, we consider the series of parameters
\be
\label{params}
\lambda_{i+1}/\lambda_i = (1 + \epsilon) \qquad \kappa_i/\sqrt{\lambda_i} = \delta  \; \; .
\ee
We define $\mu_1 = 0$ and take $\mu_2$ to be some large value (this will ensure that the values $S_i$ are small). We now try to find $\mu_{i+1}$ for $i>1$ so that
\be
\Delta t_{i}^- =  p \Delta t_i^+
\ee
for some fraction $p \in (0,1)$. For the specific example $\epsilon = \delta = 0.01$ we find numerically that the necessary $\mu_i$ values exist and form an increasing sequence for which $\mu_i/\mu_{i-1}$ converges quickly (e.g. to $\mu_i/\mu_{i-1} = 4.23$ for $p=1/2$ or $\mu_i/\mu_{i-1} = 1.067$ for $p=0.99$). The solutions corresponding to these parameters appear to satisfy our desired properties for arbitrarily large $n$, though we have not proven this analytically. Since $\lambda_n = 1.01^{n-1} \lambda_1$, we see that $\lambda_n/\lambda_1$ can be made arbitrarily large, so based on the numerical evidence we have solutions corresponding to states of CFT${}_n$ with arbitrarily small central charge that include a bubble of the spacetime dual to CFT${}_1$.

For these solutions, the regulated action is
\be
\label{eq:multiaction}
\mathcal I_{reg} = -\Delta t_1^- + \sum_{i=2}^{n-1}(1-\mu_i)(\Delta t_i^+ - \Delta t_i^-) + (1 - \mu_n) \Delta t_n^+ \; .
\ee
Defining $S_i/2$ to be the width of the $i$th CFT strip in units where the sphere radius is 1, we have
\be
\Delta t_i^+ - \Delta t_i^- = \frac{S_i}{2\sqrt{\lambda_i}} \; ,
\ee
so we can rewrite the action as
\be
\label{eq:multiaction2}
\mathcal I_{reg} = -\Delta t_1^- + \sum_{i=2}^{n}(1-\mu_i){ S_i \over 2 \sqrt{\lambda_i}}
\ee
We should compare this to the action for the other possible solutions where $\mu_1 = \cdots \mu_k = 0$ so that the first $k-1$ domain walls extend to $r=0$. For those solutions, the expression (\ref{eq:multiaction}) for the action still applies, though the parameters $\mu_{i>k}$ must be chosen to obtain the same $S_i$ values as those arising from (\ref{params}).

For the examples described above, we have checked numerically that for the other possible solutions, the $\mu$ value in each region for our original solution with only one AdS region will always be larger than the $\mu$ value in the same region for any of the other solutions.\footnote{To find the other solutions, we note that given $\mu_{k+1}$, the remaining $\mu_i$ for $i = k+2..n$ are fixed via the procedure above by demanding that $S_i$ matches with the previous solution for $i=k+1..n-1$. The value of $S_n$ for the new solution will be some function of $\mu_{k+1}$, and we must choose $\mu_{k+1}$ so that this matches with $S_n$ in the original solution.} In the case where all the $\mu$ values are large, it then follows from the form of the action (\ref{eq:multiaction2}) that the action for the solution with only one AdS region is always smallest (see right panel in Figure \ref{fig:multi_interface}).

\section{Approximating non-vacuum states}
\label{sec:excited}

\begin{figure}
    \centering
    \includegraphics[width=100mm]{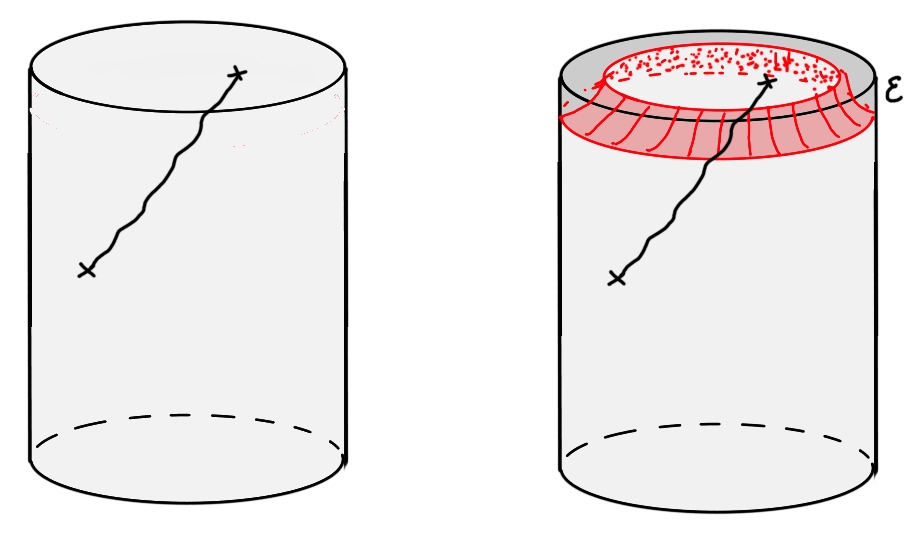}
    \caption{Approximating excited states. In the limit $\epsilon \to 0$, we expect that the boundary-to-bulk propagators on the right approach those on the left, so the interior geometry of the CFT${}_2$ state on the right should provide a good approximation to the interior geometry of the CFT state on the left.}
    \label{fig:excited}
\end{figure}

So far, we have focused on defining CFT${}_2$ states that encode a large region of the pure AdS spacetime dual to the CFT${}_1$ vacuum. However, given some other CFT${}_1$ state dual to a non-vacuum geometry, we can consider exactly the same mapping (\ref{eq:defstate}) to define a CFT${}_2$ state that approximates this. Very general perturbations to the AdS vacuum geometry can be included by adding sources for various operators to the Euclidean path integral defining the vacuum state (see, for example, \cite{Botta-Cantcheff:2015sav,Christodoulou:2016nej,Marolf:2017kvq}).\footnote{Alternatively, we can consider the insertion of a discrete set of operators.} These sources modify the boundary conditions for the corresponding bulk fields, leading to changes in the initial data. Perturbatively, these changes are governed by some Euclidean boundary-to-bulk propagator, as depicted in Figure \ref{fig:excited}. We expect that the boundary-to-bulk propagator in the background geometry dual to the approximated vacuum state (right side of figure \ref{fig:excited}) should approach the original boundary-to-bulk propagator in the limit $\epsilon \to 0$, so the interior geometry of an approximated state with sources should approach the original perturbed interior geometry in this limit.\footnote{It is apparent from the figure that for sources $\lambda_\alpha(x, \tau)$ in the original state, the best approximation to the original state at small finite $\epsilon$ may be obtained by  taking sources $\lambda_\alpha(x, \tau + \sqrt{\lambda_1} \Delta t_1)$ in the approximated case so that the bulk $\tau=0$ slice is at the same location relative to the sources in both pictures.} Assuming that our parameter choices are such that the approximated vacuum geometry with a bubble has least action, the same will be true when we add sources provided that these sources are not too large.

It would be interesting to investigate in more detail how accurately a set of bulk perturbations are reproduced in the interior of a bubble when $\epsilon$ is some small finite value.

\subsection{Approximating black hole geometries}
\label{sec:BHBH}

\begin{figure}
    \centering
    \includegraphics[width=100mm]{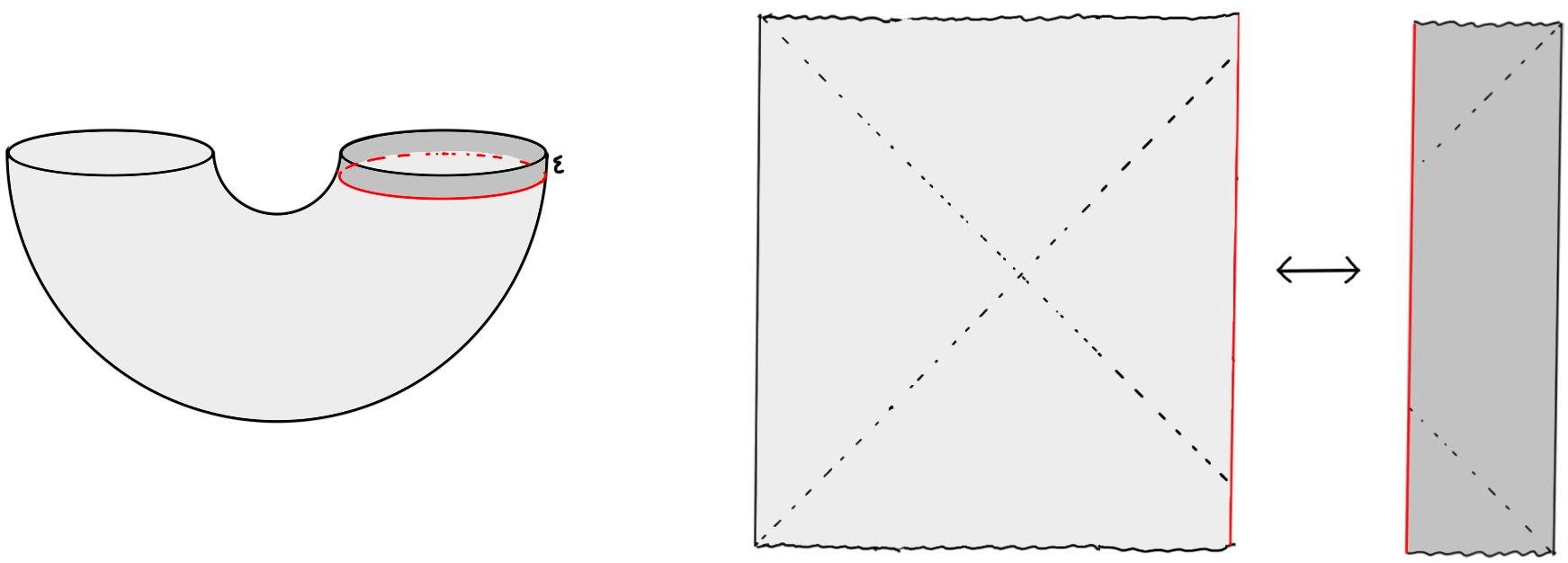}
    \caption{Approximating black hole geometries. A thermofield double state of two copies of CFT${}_1$ is replaced by an entangled state of CFT${}_1$ and CFT${}_2$. The dual geometry is a Lorentzian wormhole connecting regions described asymptotically by the low-energy gravitational theories dual to CFT${}_1$ and CFT${}_2$, with an interface brane between them.}
    \label{fig:tfd}
\end{figure}

We can also consider the case of highly-excited CFT states, where the bulk geometry is far from pure AdS. In the generic situation, we expect that the bulk geometry should be well-approximated by an AdS/Schwarzschild black hole, so we will consider the case where CFT${}_1$ is in a thermal state, purified as the thermofield double state of two copies of CFT${}_1$. In this case, we can approximate the state as an entangled state of CFT${}_1$ and CFT${}_2$, defined by\footnote{Alternatively, we could have considered replacing both sides with CFT${}_2$.}
\beas
|\Psi \rangle &=& \identity \otimes M_{{\cal I},\epsilon} |\Psi_{TFD} \rangle_{LR} \cr
&=& {1 \over Z^{1 \over 2}} \sum_i e^{-\frac{\beta}{2}E_i} |E_i \rangle_L \otimes M_{{\cal I},\epsilon} |E_i \rangle_R
\eeas
The state arises from the Euclidean path integral shown at the left in Figure \ref{fig:tfd}.

In this case, the dual geometry (either in the Euclidean or in the Lorentzian picture) will be one where both parts are portions of Schwarzschild-AdS geometries with different values of $\mu$. One possibility for the Lorentzian geometry is shown at the right in Figure \ref{fig:tfd}, though we can also have the situation that the domain wall is inside the horizon on the right. Similar geometries have been discussed in \cite{Barbon:2010gn}.

For these black hole geometries the interface trajectories are determined in the same way as before. The explicit results for $\Delta t_1$ and $\Delta t_2$ in this general case with nonzero $\mu_1$ and $\mu_2$ are given in appendix \ref{app:interface}.

\subsubsection*{Phases}

A basic way to characterize the solutions corresponding to entangled states of CFT${}_1$ and CFT${}_2$ is to understand whether the interface brane lies behind the horizon on the CFT${}_1$ side, the CFT${}_2$ side, or both sides. Since the two CFTs have no interaction, the interface must be behind the horizon on at least one side, otherwise we could send a signal from one asymptotic region to the other.

\begin{figure}
    \centering
    \includegraphics[width=150mm]{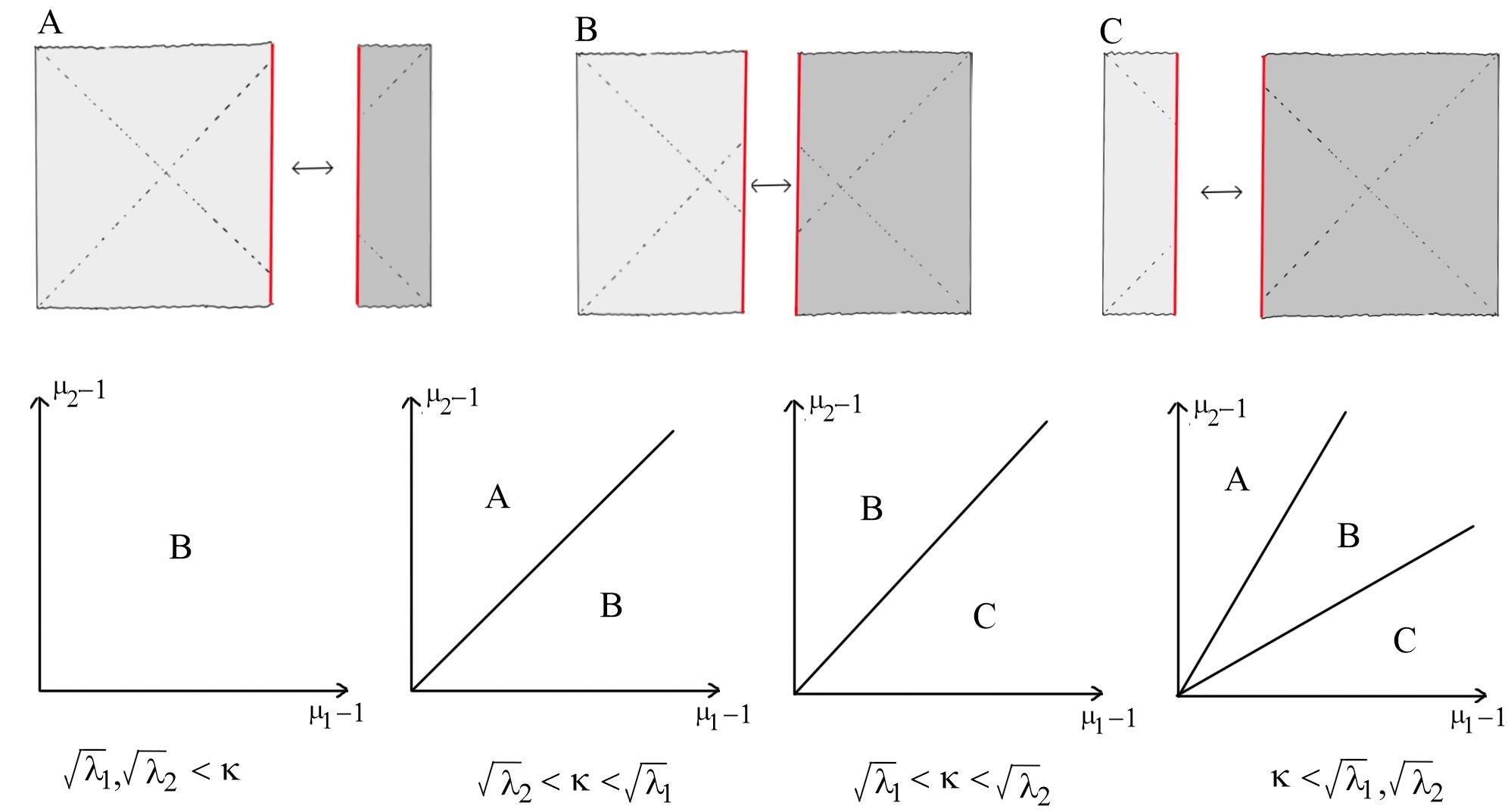}
    \caption{Phases for black hole geometries dual to entangled states of CFT${}_1$ and CFT${}_2$.}
    \label{fig:BHphases}
\end{figure}

The interface brane will be behind the horizon on the CFT${}_1$ side (as in F
Figure \ref{fig:tfd}) if and only if $\dot{t}_1(r_0) > 0$, while the interface brane will be behind the horizon on the CFT${}_2$ side if and only if $\dot{t}_2(r_0) < 0$.  From equation (\ref{tpmeq}), these give
\bea
(\lambda_1 - \lambda_2 + \kappa^2)r_0^2 + (\mu_2 - \mu_1) > 0 \qquad {\rm interface \; behind \; CFT_1 \; horizon}\cr
(\lambda_1 - \lambda_2 - \kappa^2)r_0^2 + (\mu_2 - \mu_1) < 0 \qquad {\rm interface \; behind \; CFT_2 \; horizon}
\label{BHboundaries}
\eea
The phase boundaries lie at $\dot{t}_1(r_0) = 0$ and $\dot{t}_2(r_0) = 0$; when these are satisfied, we also have $r_0 = r_H^1$ and $r_0 = r_H^2$ respectively, since the phase boundary is the boundary in parameter space where the interface crosses the Euclidean horizon. Using (\ref{eq:defrH}) in (\ref{BHboundaries}), the phase boundaries simplify to
\bea
\lambda_1 (\mu_2 - 1) > (\lambda_2 - \kappa^2)(\mu_1 - 1) \qquad {\rm interface \; behind \; CFT_1 \; horizon}\cr
\lambda_2 (\mu_1 - 1) > (\lambda_1 - \kappa^2)(\mu_2 - 1) \qquad {\rm interface \; behind \; CFT_2 \; horizon}
\label{BHboundaries2}
\eea
Figure \ref{fig:BHphases} shows the types of black hole geometries that are possible and for which parameter values each phase is realized.  The slopes of the various lines can be read off from (\ref{BHboundaries2}).

Generally, we find that configurations similar to that shown in Figure \ref{fig:tfd} with CFT${}_2$ occupying a small fraction of the thermal circle corresponds to parameter values $\mu_2 \gg \mu_1$.

\section{Properties of the CFT state}
\label{sec:CFTstates}

It is interesting to consider the properties of the CFT states which are able to encode a bubble of pure AdS spacetime with size $r_0$. In particular, we can ask what is the minimum amount of energy needed in CFT${}_2$ to describe a state that includes a bubble of size $r_0$ for the AdS spacetime dual to CFT${}_1$.

For $D=3$, the energy and (coarse-grained) entropy of the CFT${}_2$ state  can be related to the parameter $\mu$ as\footnote{Here, the entropy can be derived using the Bekenstein-Hawking formula $S = A_H/(4G)$ while the energy can be obtained using $dE = TdS$, taking $E=0$ for $\mu = 0$ so $E_{CFT}$ is the energy relative to the vacuum. The temperature is the inverse periodicity of the CFT time, $T^{-1} = \beta_{CFT} = \sqrt{\lambda} \beta_{grav}$, where $\beta_{grav}$ is given in (\ref{defbeta}).}
\bea
E_{CFT} &=& {c_2 \over 12} \mu \cr
S_{CFT} &=& {c_2  \over 3} \pi \sqrt{\mu -1}
\label{ESCFT}
\eea
where we have that
\bea
\label{eq:mumin}
\mu &=& 2 \kappa r_0 (1 + \lambda_1 r_0^2)^{1 \over 2} + r_0^2 (\lambda_2 - \lambda_1 - \kappa^2) \cr
&=& r_0^2[\lambda_2 - (\sqrt{\lambda_1} - \kappa)^2] + {\kappa \over \sqrt{\lambda_1}} + {\cal O}(r_0^{-2}) \; .
\eea
For a given large value of $r_0$, $\mu$ is minimized for $\kappa \to \sqrt{\lambda_1} + \sqrt{\lambda_2}$ (corresponding to $\log g \to \infty)$ when $\lambda_2 > \lambda_1$ and for $\kappa \to \sqrt{\lambda_1} - \sqrt{\lambda_2}$ (corresponding to $\log g \to -\infty)$ when $\lambda_2 < \lambda_1$. Thus, the least energy will be achieved by choosing the interface with the largest value of $\log g$ when $c_2 > c_1$ and the smallest value of $\log g$ when $c_2 < c_1$.

In terms of CFT parameters, we have that the energy of a CFT${}_2$ state required to describe a bubble of size $r_0 = n L_1$  of the AdS spacetime dual to the CFT${}_1$ vacuum (i.e. for a bubble whose radius is  $n$ AdS lengths) satisfies
\be
E >  \left\{\ba{ll} {1 \over 12}(c_2 + c_1)(1 - {1 \over 4 n^2} + {\cal O}(n^{-4})) & \qquad c_2 < c_1 \cr
{1 \over 12}(c_2 - c_1)(1 - {1 \over 4 n^2} + {\cal O}(n^{-4})) & \qquad c_2 > c_1 \ea \right.
\ee
We recall that $c_2/12$ is the energy of the CFT state associated with the lightest BTZ black hole. Thus, when approximating the CFT${}_1$ vacuum with a CFT with less degrees of freedom the state necessarily has energy in the black hole range, while for a CFT${}_2$ with more degrees of freedom, a lighter state is possible. While these infimum values for the energies (which require taking $\log g \to \pm \infty$) have a finite limit for $r_0 \to \infty$, we see that for a fixed choice of CFT parameters, the energy of a CFT${}_2$ state approximating an AdS bubble of radius $r_0$ always scales as $r_0^2$ for large $r_0$.

\subsubsection*{Energy of approximated black hole states}

When approximating a bubble of radius $r_0$ of a black hole geometry with mass parameter $\mu_1$, we find that the equation (\ref{eq:mumin}) for the parameter $\mu$ that determines the energy and entropy of the CFT${}_2$ state via (\ref{ESCFT}) generalizes to
\beas
\mu_2 &=& \mu_1 +  2 \kappa r_0 (1 + \lambda_1 r_0^2 - \mu_1)^{1 \over 2} + r_0^2 (\lambda_2 - \lambda_1 - \kappa^2) \cr
&=& r_0^2[\lambda_2 - (\sqrt{\lambda_1} - \kappa)^2] + {\kappa \over \sqrt{\lambda_1}} + \mu_1 \left(1 - {\kappa \over \sqrt{\lambda_1}} \right) + {\cal O}(r_0^{-2}) \; .
\eeas
where we are assuming that $\dot{t}_1(r_0) > 0$ so that bubble at $t=0$ contains the black hole as in Figure \ref{fig:BHphases}A,B. An interesting feature is that for $\kappa > \sqrt{\lambda_1}$ the CFT${}_2$ state energy actually decreases as the black hole mass parameter $\mu_1$ increases, though this dependence is only in the subleading ${\cal O}(r_0^0)$ behavior. So, for fixed $r_0$, it does not necessarily require more energy in CFT${}_2$ to approximate an excited state of CFT${}_1$.

\section{Discussion}
\label{sec:discussion}

In the context of a simple holographic model that captures the central charges of CFTs and the interface entropy of possible interfaces between them, we have found that it is always possible to find states of CFT${}_2$ whose dual geometries include arbitrarily large causal patches of the geometry dual to the vacuum state of CFT${}_1$ or a very general class of excited states.
Our result should hold generally provided that there is always some choice of interface (or series of interfaces) and some sufficiently small $S$ for which the topology of the interface brane takes the form shown in Figure \ref{fig:Euclidean}c rather than the form shown in Figure \ref{fig:Euclidean}b. It would be interesting to see whether this is true in more general models, for example, in models with more general bottom-up actions or microscopic models arising from string theory.

In our simple model, the interface brane is assumed to couple only gravitationally. Combined with the assumed spherical symmetry this implies that the interior geometry of our bubbles is precisely pure AdS with AdS length $L_1$, i.e. the geometry of the bubble interior for the approximated state is precisely the same as the geometry that we started with. In more general cases, the interface brane can couple to other fields in the geometry, so it is interesting to ask whether the bubble interior would still be approximately pure AdS in these cases. The following argument suggests that this should be the case.

Suppose that the interior geometry at $t = 0$ contained some massive particle. In the Euclidean geometry, this particle will lie on a time-symmetric geodesic. If this geodesic is not too close to the bubble wall, it will intersect the asymptotically AdS boundary at points in the past and the future in the asymptotic region associated with CFT${}_1$. But such an intersection would imply the insertion of a heavy operator in the CFT, and we have assumed that there aren't any such insertions. Similarly, we expect that extra light matter in the bubble interior would imply a modification of the asymptotic behavior of fields in the CFT${}_1$ region of the asymptotic geometry, corresponding to sources for various light operators in the CFT. Assuming that these sources are not present, we expect that the bubble interior away from the bubble wall should be very close to vacuum AdS. It would be interesting to investigate this further, for example in a model where the interface brane action includes a source for a scalar field.

\subsubsection*{Probing the bubble interior with CFT${}_2$}

In cases where the dual geometry to our CFT${}_2$ includes a bubble of the AdS spacetime dual to the CFT${}_1$ vacuum or to perturbative excitations around this geometry, it is interesting to ask which CFT${}_2$ observables can be used to probe the physics inside the bubble. In simple cases, where the bubble wall is not always behind a black hole horizon, we expect that an HKLL-type construction \cite{Hamilton:2005ju,Hamilton:2006az} should work to express interior bulk operators in terms of boundary operators. Here, we need a propagator from the interior of the bubble to the asymptotically AdS region outside the bubble; this will generally involve different fields on either side of the bubble, and will require detailed knowledge of the physics of the interface brane to understand how excitations on one side will be transmitted to the other side.

Calculations of entanglement entropy for various spatial regions of CFT${}_2$ should also probe the bubble interior in some cases, even when the bubble lies behind the black hole horizon. Calculations similar to those in \cite{Cooper2018} could be used to investigate which portion of the interior region is reached by RT surfaces (for various values of the parameters) and in which cases some or all of this region lies in the ``entanglement shadow''.

In general, we expect that the local description of physics inside the bubble should be quite complicated from the CFT${}_2$ perspective when the bubble lies behind the black hole horizon. Even without an interface, understanding the description of local physics behind a black hole horizon is a significant open question.

\subsubsection*{Relation to ER=EPR and models of black hole evaporation}

The geometries of section \ref{sec:BHBH} provide an explicit realization of the suggestion in \cite{VanRaamsdonk:2009ar} that certain entangled states of different holographic CFTs should correspond to wormhole geometries with asymptotic regions separated by a dynamical interface, allowing entanglement to connect spacetime regions even when described by different low-energy gravitational theories.

These explicit examples may help clarify a puzzling feature of the ER=EPR story \cite{Maldacena2013} and its recent manifestation in examples of evaporating black holes  described by a conventional holographic theory coupled to an auxiliary radiation system \cite{GP,AE,Almheiri:2019yqk}. There, degrees of freedom describing the Hawking radiation of a black hole are entangled with degrees of freedom describing the remaining black hole. For an old black hole, the radiation degrees of freedom are proposed to encode the physics behind the black hole horizon, so that we have some type of wormhole between the black hole exterior and another spacetime region described by the radiation system. But it may seem puzzling how this can be possible if the radiation system is described by holographic degrees of freedom dual to some other theory of quantum gravity, or degrees of freedom which are not holographic. But our examples show that this puzzle can be resolved provided the full geometry has some interface between the two theories of gravity, or between the ordinary gravitational theory describing the black hole and some ``dual'' of the non-holographic degrees of freedom which isn't a conventional theory of gravity.

\subsubsection*{General picture}

To conclude, we summarize the speculative picture that has motivated our study and that we have found evidence for using our simple holographic model. First, we have suggested that:

\begin{itemize}
    \item
    Theories of gravity dual to CFTs that can be non-trivially coupled at an interface are part of the same non-perturbative theory. This includes dynamical interface branes that can connect regions of spacetime described by the low-energy gravitational theories associated with the individual CFTs.
\end{itemize}
According to this, distinct non-perturbative theories of quantum gravity would correspond to different equivalence classes of CFTs, where two CFTs are in the same equivalence class if they can be coupled non-trivially at an interface. An interesting possibility is that there is only one equivalence class (or at least, only one containing holographic theories), in which case all low-energy gravitational theories (with asymptotically AdS vacua) would be part of a single non-perturbative theory of gravity, as advocated recently in \cite{VanRaamsdonk:2018zws,McNamara:2019rup}.

The idea that many apparently different gravitational theories may all be related is familiar from string theory,\footnote{That all consistent gravitational theories should come from string theory and be related to each other has been advocated recently in \cite{Kim:2019ths}.} and is central to the multiverse picture of cosmology (see e.g. \cite{Bousso:2000xa}), in which bubbles with physics described by one low-energy effective theory may form and expand in an ambient spacetime described by another low-energy effective theory. However, suppose there were some other theory of quantum gravity, apparently unrelated to string theory, describing physics in an asymptotically AdS spacetime. According to the general principles of AdS/CFT, the asymptotic observables of this theory should define a dual CFT (see e.g. \cite{Marolf:2008mf}). Provided that this CFT can be nontrivially coupled at an interface to some CFT dual to a gravitational theory arising from string theory, then the gravitational physics of the interface theory should involve both low-energy gravitational theories in different regions separated by a dynamical interface, and thus be part of a single non-perturbative theory.

Our main focus has been to provide evidence for the following:

\begin{itemize}
    \item
    Given $|\Psi \rangle_{CFT_1}$ dual to an asymptotically AdS spacetime, and given CFT${}_2$ that can be nontrivially coupled to CFT${}_1$ at an interface, we can find a state $|\Psi \rangle_{CFT_2}$ whose dual geometry includes an arbitrarily large portion of the Wheeler-DeWitt patch associated with $|\Psi \rangle_{CFT_1}$
\end{itemize}
We have given an explicit construction of the states $|\Psi \rangle_{CFT_2}$ in terms of a quench operator defined using a Euclidean path integral for the interface CFT.

We emphasize that the interior regions of the bubbles described by these states of CFT${}_2$ include all the usual low-energy fields of the gravitational theory associated with CFT${}_1$, even if CFT${}_2$ and CFT${}_1$ have very different operator spectra.

According to this general picture, different holographic CFTs correspond to different possible asymptotically AdS behaviors for quantum gravity, while the gravitational physics of finite patches of spacetime is something that can be encoded in any CFT in the same equivalence class as the original one. That this same physics can be described by many different holographic theories suggests that the gravitational physics of such finite-volume patches of spacetime may be dual to some more universal physics that is present in the IR physics of any holographic CFT (such as the evolution of entanglement structure or complexity \cite{Brown:2015lvg} of the states).\footnote{An observation of Alex May is that this universality may be closely related to universality in computing. We can interpret the results as saying that different CFTs in the same class have the same computational power in simulating quantum gravity on finite-volume spacetime regions.} In fact, our CFT construction of states applies just as well for non-holographic CFTs (i.e. where the ``gravity'' description would not be in terms of a conventional low-energy effective theory of gravity), so an interesting possibility is that the precise physics of quantum gravity could in principle be simulated through the IR quantum dynamics of an ordinary material, though setting up an appropriate state and understanding the proper observables to measure in order to extract interesting physics is clearly a significant challenge.

\section*{Acknowledgements}

We would like to thank Alex May and the rest of the UBC string group, as well as Raphael Bousso and the the group at Berkeley BCTP for useful comments and discussion. We would like to thank the organizers of the QIQG5 conference at Davis, where this project originated.  MVR is supported by the Simons Foundation via the It From Qubit Collaboration and a Simons Investigator Award. PS is supported by the Natural Sciences and Engineering Research Council of Canada via the Canada Graduate Scholarship.

\appendix

\section{Solving the junction conditions}
\label{appendix:junctions}

We will derive the expression
\be\label{JC2_Schwarzschild_appendix}
f_1 {dt_1 \over ds} + f_2 {dt_2 \over ds} = \kappa r
\ee
starting from the second junction condition
\begin{equation}\label{JC2_appendix}
    K_{1ab}-K_{2ab}=\kappa h_{ab}.
\end{equation}
First, we denote the proper velocity of a particle in the domain wall in the coordinates $(t_i,r)$ as $u^a=(\dot t_i,\dot r)=(dt_i/ds,dr/ds)$, where we are suppressing the angular components which vanish by spherical symmetry. In Euclidean signature $u^a u_a=1$ and thus
\be
f_i \left({dt_i \over ds}\right)^2 + {1 \over f_i}  \left({dr \over ds}\right)^2 = 1 \; .
\ee
The normal to the domain wall, $n^a$, satisfies $u^a n_a=0$ and $n^a n_a = 1$. We can easily confirm that $n_a = (-\dot r,\dot t_i)$ fulfills these requirements. Our choice of sign for $n_a$ is consistent with our convention that this vector points from region 1 into region 2.

Next let us compute the $K_{1\theta\theta}=n_{\theta;\theta}$. Since $n_\theta=0$ we have
\begin{equation}
    K_{1\theta\theta} = -\Gamma^i_{\theta\theta}n_i=
-\Gamma^r_{\theta\theta}n_r=r f_1(r) \frac{d t_1}{d s}
\end{equation}
Similarly $K_{2\theta\theta}=-n_{\theta;\theta}$ gives
\begin{equation}
    K_{1\theta\theta} = -r f_2(r) \frac{t_2}{d s},
\end{equation}
where the minus sign arises because the extrinsic curvature is defined using an outward facing normal. Substituting these expressions into the $\theta\theta$ component of (\ref{JC2_appendix}) gives (\ref{JC2_Schwarzschild_appendix}), as desired. Note that all other angular components of (\ref{JC2_appendix}) lead to the same equation.

\section{Interface trajectories in $D=3$}
\label{app:interface}

For an interface of tension $\kappa$ between a region with parameters $\lambda_1$, $\mu_1$ and parameters $\lambda_2, \mu_2$, the interface trajectory is determined by
\be
\left({dr \over ds}\right)^2 = V_{eff}
\ee
with
\be
V_{eff} = 1 + \lambda_1 r^2 - \mu_1 - \left[ {(\lambda_2 - \lambda_1 - \kappa^2) \over 2 \kappa} r - {\mu_2 - \mu_1 \over 2 \kappa r}\right]^2 \; .
\ee
The interface reaches a minimum value $r_0$ where $V_{eff}(r_0) = 0$. We have
\be
r_0 = \frac{|\Delta \mu|}{\sqrt{2 \kappa^2 (1-\mu_1)\!+\!(\Delta \lambda-\kappa^2)(\Delta \mu)\!+\!2\sqrt{(1-\mu_1)^2\kappa^4 +(1-\mu_1)(\Delta \lambda-\kappa^2)\kappa^2 \Delta \mu+\kappa^2\lambda_1 \Delta \mu^2}}}
\ee
where $\Delta \mu = \mu_2 - \mu_1$ and $\Delta \lambda = \lambda_2 - \lambda_1$.

The trajectories in the $\lambda_1$ and $\lambda_2$ regions are respectively
\bea
 {dt_1 \over dr} &=& {1 \over f_1 \sqrt{V_{eff}}} \left({1 \over 2 \kappa r}(f_1 - f_2) + {1 \over 2} \kappa r\right) \cr
 {dt_2 \over dr} &=& -{1 \over f_2 \sqrt{V_{eff}}} \left({1 \over 2 \kappa r}(f_2 - f_1) + {1 \over 2} \kappa r\right) \; ,
\eea
where $f_i=1+\lambda_i r^2-\mu_i$. We find that
\be\label{eq:BHexact_appendix}
\Delta t_2 \equiv t_2(\infty) - t_2(r_0) =  \Delta(\lambda_1,\lambda_2,\mu_1,\mu_2, \kappa) \qquad \Delta t_1 \equiv t_1(\infty) - t_1(r_0)= -\Delta(\lambda_2,\lambda_1,\mu_2,\mu_1, \kappa)
\ee
where
\begin{equation}
\Delta(\lambda_1,\lambda_2,\mu_1,\mu_2,\kappa) = C (K(z) + C_\Pi \Pi(\nu,z)) \; .
\label{dtgen}
\end{equation}
Here, $\Pi$ and $K$ are elliptic integral functions
defined as
\begin{equation}
    \Pi(\nu,z) = \int_0^1 {dt \over (1 - \nu t^2) \sqrt{(1 - t^2)(1 - z^2 t^2)}} \qquad K(z) = \Pi(0,z)
\end{equation}
and the the parameters in (\ref{dtgen}) are defined as
\begin{eqnarray}
z &=& \sqrt{(\mu_2 - \mu_1)^2 \over (\mu_2 - \mu_1)^2 + 4 r_0^4 \kappa^2 A} \qquad
\nu = {1 - \mu_2 \over 1 - \mu_2 + \lambda_2 r_0^2} \cr
C &=&   z r_0 {1 \over 1 - \mu_2} {\rm sign}(\mu_2 - \mu_1) \qquad
C_\Pi = {[(1 - \mu_1)\lambda_2 + (1 - \mu_2) (\kappa^2 - \lambda_1)]r_0^2 \nu \over (1 - \mu_2)(\mu_1 - \mu_2)}
\end{eqnarray}
The parameter $A$ is defined in (\ref{defA}).

For $\mu_1 = 0$ and $\mu_2=\mu$ these results reduce to those in section \ref{sec:BHsolutions}.

\subsection{Large $\mu$}
\label{app:smallS}

Continuing to take $\mu_1 = 0$ and $\mu_2=\mu$, starting from equation (\ref{eq:BHexact_appendix}) we find that in the limit of large $\mu$, we have
\begin{equation}
\label{eq:dt1app}
\Delta t_1 = \sqrt{\frac{\kappa}{\mu}} \left(K(z_\infty) - {\pi \over 2}  z_\infty^2 \; {}_2 F_1\left({1 \over 2}, {3 \over 2} ; 2|z_\infty^2\right) \right)
\end{equation}
and
\begin{equation}
\Delta t_2 = -{1 \over \sqrt{\kappa \mu}} \left(K(z_\infty) + {\kappa+1 \over \kappa - 1} \Pi \left(1 - {\lambda_2 \over (\kappa-1)^2}, z_\infty \right) \right)
\end{equation}
where
\begin{equation}
z_{\infty} = \sqrt{\lambda_2 - (\kappa-1)^2 \over 4 \kappa} \; .
\end{equation}
Here, we have used that for small $\nu$,
\begin{equation}
    \Pi(\nu,z) = K(z) + {\pi \over 4} \nu \; {}_2 F_1\left({1 \over 2}, {3 \over 2} ; 2|z^2\right)  + {\cal O}(\nu^2)
\end{equation}

\section{Comparing on shell actions}
\label{app:actions}

For some CFT parameter values there are multiple bulk geometries consistent with the given CFT data. In this case, the geometry that describes the CFT in the semiclassical limit is the one with the lowest value of the Euclidean action.

Recall that the Euclidean action for a domain wall solution is given by:
\begin{align}
    \mathcal I=-\frac{1}{16\pi G_D}
    \Bigg[&
    \int_{\mathcal{M}_1}\d[D]{x}\sqrt{g_1}\left(R_1-2\Lambda_1\right)
    +\int_{\mathcal{M}_2}\d[D]{x}\sqrt{g_2}\left(R_2-2\Lambda_2\right)\notag\\
    &
    +
    2\int_\mathcal{S}\d[D-1]{y}\sqrt{h}(K_1-K_2)
    -2(D-2)\int_\mathcal{S}\d[D-1]{y}\sqrt{h} \kappa
    \Bigg],
\end{align}
where we have used $\kappa\equiv 8\pi G_D T/(D-2)$. Taking the trace of the Einstein equation gives
\begin{equation}
    R_i = \frac{2D}{D-2}\Lambda_i,
\end{equation}
for $i\in\{1,2\}$, and taking the trace of the second junction condition, Eq.~\eqref{eq:JC2} gives
\begin{equation}
    K_1-K_2=(D-1)\kappa.
\end{equation}
The on shell Euclidean action thus simplifies to
\begin{align}\label{eq:On_shell_Euclidean_action}
    \mathcal I=-\frac{1}{8\pi G_D}
    \left[
    \frac{2\Lambda_1}{D-2}
    \int_{\mathcal{M}_1}\d[D]{x}\sqrt{g_1}
    +
    \frac{2\Lambda_2}{D-2}
    \int_{\mathcal{M}_2}\d[D]{x}\sqrt{g_2}
    +\kappa\int_\mathcal{S}\d[D-1]{y}\sqrt{h}
    \right].
\end{align}

The volume element in region $\mathcal M_1$ is
\begin{equation}\label{eq:volume_element_Mi}
    d^D x\sqrt{g_i} =
    r^{D-2}dt_i dr d\Omega_{D-2},
\end{equation}
where $d\Omega_{D-2}$ is the volume element on $S^{D-2}$. Parametrizing the domain wall by $t_i=t_i(r)$, we find that the volume element on the domain wall is given by
\begin{equation}
    d^{D-1}y\sqrt{h} =
    r^{D-2}\sqrt{f_i(r)\left(\frac{dt_i}{dr}\right)^2+\frac{1}{f_i(r)}} dr d\Omega_{D-2}.
\end{equation}
Note that the proper time $\tau$ for a particle on the domain wall at constant angular coordinates is given by $d\tau^2 = f_i(r)dt_i^2+dr^2/f_i(r)$. Combining this with the equation of motion \eqref{req} of the domain wall gives
\begin{equation}
    \left(\frac{dt_i}{dr}\right)^2
    =
    \frac{\alpha_i^2(r)}{f_i^2(r)V_\text{eff}(r)},
\end{equation}
and hence the volume element on the domain wall $\mathcal S$ is
\begin{equation}\label{eq:Volume_element_S}
    d^{D-1}y\sqrt{h} =
    \frac{r^{D-2}}{\sqrt{V_\text{eff}(r)}} dr d\Omega_{D-2}.
\end{equation}
Using Eqs.~\eqref{eq:volume_element_Mi} and \eqref{eq:Volume_element_S} in the on shell Euclidean action \eqref{eq:On_shell_Euclidean_action} and performing the integral over $S^{D-2}$ gives
\begin{align}\label{eq:On_shell_Euclidean_action_simplified}
    \mathcal I=-\frac{\Vol\left(S^{D-2}\right)}{8\pi G_D}
    \left[
    \frac{2\Lambda_1}{D-2}
    \int_{\mathcal{M}_1}r^{D-2} dt_1 dr
    +
    \frac{2\Lambda_2}{D-2}
    \int_{\mathcal{M}_2}r^{D-2} dt_2 dr
    +\kappa\int_\mathcal{S}\frac{r^{D-2}}{\sqrt{V_\text{eff}(r)}} dr
    \right],
\end{align}
where $\Vol\left(S^{D-2}\right)$ is the volume of $S^{D-2}$. Note that the integrals $\int_{\mathcal M_i}$ are now just in the $t_i-r$ plane, while the integral $\int_{\mathcal S}$ is just over $r$. The bounds on these integrals depend on which domain wall geometry we are considering.

\subsection{Simplifications in D = 3}

Starting from the expression (\ref{eq:On_shell_Euclidean_action}) for the action and specializing to $D=3$, we have that
\begin{eqnarray}
    4 G_3 \mathcal{I} &=&  \int_{\mathcal{M}_1} d \omega_1 +  \int_{\mathcal{M}_2} d \omega_2 - \kappa \int_\mathcal{S} r d \tau \cr
    &=& \int_{\partial \mathcal{M}_1} \omega_1 +  \int_{\partial \mathcal{M}_2} \omega_2 - \kappa \int_\mathcal{S} r d \tau
    \label{Form_action}
\end{eqnarray}
where we are now considering $\mathcal{M}_i$ as regions on the two-dimensional $r-t_i$ plane and we have defined the one-forms
\begin{equation}
    \omega_i = \lambda_i r^2 d t_i \; .
\end{equation}

The non-vanishing contributions to the integral are those corresponding to the cutoff surface, the domain wall, and (in cases $IIA-$ and $IIB-$) the horizon $r = r_\textsc{h}$. In all cases, the contribution from the cutoff surfaces in $\mathcal{M}_1$ and $\mathcal{M}_2$ give
\begin{equation}
    4G_3\mathcal{I}_{cutoff} = r_{max}^2 t_{max}
\end{equation}
up to corrections which vanish in the limit $r_{max} \to \infty$.\footnote{Here, we are taking the asymptotic time coordinate to run from 0 to $S/2$ in the $\lambda = \lambda_1$ region and $S/2$ to $t_{max}$ in the $\lambda = \lambda_2$ region. The result including corrections is $r_{max}^2 t_{max} - t_2(r_{max}) + t_1(r_{max})$.}

To express the contribution from the domain wall $\mathcal S$, we can combine the three contributions at the interface into the single integral
\begin{equation}
    4G_3\mathcal{I}_{\mathcal S} = \int_\mathcal{S} (\lambda_1 r^2 dt_1 - \lambda_2 r^2 dt_2 - \kappa r d \tau).
\end{equation}
Here we are integrating along the direction of increasing $r$, and so the sign of the second term is changed relative to Eq.~\eqref{Form_action}, which required integrating in the opposite direction. To simplify this, we note that $\alpha_i = f_i dt_i/d \tau$, so the second junction condition can be written as
\begin{equation}
    f_1 dt_1 - f_2 dt_2 = \kappa r d \tau
\end{equation}
where $\tau$ is taken to increase in the direction of increasing $r$. Using $f_i = 1 - \mu_i + \lambda_i r^2$, we have that
\begin{equation}\label{eq:Action_3d_full_simplified}
    4G_3\mathcal{I}_{\mathcal S} = \int_\mathcal{S} (- dt_1 + (1-\mu) dt_2) = (1 - \mu)\Delta t_2 - \Delta t_1 .
\end{equation}
Here, $\Delta t$ represents the change in $t$ from the minimum $r$ location on the interface to the asymptotic boundary.\footnote{More precisely, it is the change going to the cutoff surface, but the difference goes to zero when the cutoff surface is taken to infinity.}

Finally, in cases where the interior regions includes a Euclidean horizon, we have a contribution
\begin{equation}
    4G_3\mathcal{I}_{horizon} = - \lambda_2 r_\textsc{h}^2 \beta/2 = (1 - \mu) \beta/2.
\end{equation}
Here and above, we are calculating the action only in the region $t \ge 0$. The action in region $t\le 0$ is equal by symmetry.

In summary, for AdS geometries, the regulated action is
\begin{equation}
\label{acAdS_Appendix}
    4G_3\mathcal{I}^{reg}_{AdS} = \Delta t_2 - \Delta t_1
\end{equation}
while for black hole geometries, the regulated action is
\begin{equation}
\label{acBH_Appendix}
    4G_3\mathcal{I}^{reg}_{BH} = (1 - \mu) {S \over 2\sqrt{\lambda_2}} - \Delta t_1
\end{equation}
In this expression, we have used that $S/2 = \sqrt{\lambda_2} \Delta t_2$ in black hole exterior solutions with no horizon, while $S/2 = \sqrt{\lambda_2}(\Delta t_2 + \beta/2)$ in exterior solutions with a horizon.

\bibliographystyle{jhep}
\bibliography{references}

\end{document}